\documentclass[reprint,aps,pra,amsmath,amssymb,floatfix]{revtex4-2}
\usepackage{revquantum}
\usepackage{color}
\usepackage{import}
\usepackage{graphicx}
\usepackage{verbatim}
\graphicspath{{img/}}

\def\fa{\ensuremath{\rotatebox[origin=c]{90}{\footnotesize $\circlearrowleft$}}}
\def\fc{\ensuremath{\rotatebox[origin=c]{270}{\footnotesize $\circlearrowright$}}}

\definecolor{grey}{rgb}{0.62,0.77,1}
\definecolor{brown}{rgb}{0.67,0.47,0}
\newcommand{\abs}[1]{\ensuremath{\vert#1\vert}}

\newcommand{\EI}{\text{A}}
\newcommand{\EII}{\text{B}}
\renewcommand{\L}{L_{\EI\EII}}
\newcommand{\LL}{L_{\EI D}}

\begin{document}
\title{Dissipation-enabled resonant adiabatic quantum state transfer: \\
Entanglement generation and quantum cloning}

\author{Marvin Gajewski}
\thanks{Present adress: Theoretische Physik, Universit\"at des Saarlandes, 66123 Saarbr\"ucken, Germany}
\email{marvin.gajewski@uni-saarland.de}
\affiliation{Institut f\"ur Angewandte Physik, Technische Universit\"at Darmstadt, 64289 Darmstadt, Germany}
\author{Thorsten Haase}
\affiliation{Institut f\"ur Angewandte Physik, Technische Universit\"at Darmstadt, 64289 Darmstadt, Germany}
\author{Gernot Alber}
\affiliation{Institut f\"ur Angewandte Physik, Technische Universit\"at Darmstadt, 64289 Darmstadt, Germany}

\begin{abstract}
    Resonant dissipation-enabled adiabatic quantum state transfer processes between the polarization degrees of freedom of a single-photon wave packet and quantum emitters are discussed. These investigations generalize previous work [N.~Trautmann  and G.~Alber, Phys.  Rev. A \textbf{93}, 053807 (2016)] 
    by taking into account the properties of the spontaneously emitted photon wave packet and of non adiabatic corrections. It is demonstrated that the photonic degrees of freedoms of these adiabatic one-photon quantum state transfer processes can be used for the passive, heralded, and deterministic preparation of Bell states of two material quantum emitters and for realizing a large family of symmetric and asymmetric quantum cloning processes. Although these theoretical investigations concentrate on waveguide scenarios they are expected to be relevant also for other scenarios as long as the processes involved are adiabatic so that the Fourier-limited bandwidth of the single-photon wave packet involved is small in comparison with the relevant dissipative rates.
\end{abstract}

\keywords{Quantum optics, Entanglement, Quantum cloning}

\maketitle

\section{Introduction}
Realizing highly efficient quantum state transfer processes from the polarization degrees of freedom of a single
photon wave packet to material quantum systems 
is crucial for advancing quantum technology. Their
potential applications are particularly relevant for advancing quantum communication with the ultimate aim of realizing a quantum internet \cite{Kimble2008QuantumInternet}.
Numerous proposals and realizations for interfacing stationary and flying qubits as well as performing information processing have already been made in the framework of cavity quantum electrodynamics (QED)  \cite{Cirac1997NetworkNodeCavity, Stute2012NetworkNodeIonCavity,Rempe2012ElementaryNode,Rempe2020}, of solid-state platforms \cite{Togan2010NetworkNodeSolidState, Bussieres2014NetworkNodeSolidState}, of photonic nanostructures \cite{Lodahl2015NetworkNodePhotonicNanostructure, Nguyen2019NetworkNodeSolidStateCavity} and of superconducting platforms \cite{Flurin2015NetworkNodeSuperconducting}.
In this context passive interfaces which 
do not require external control or feedback and, hence, act autonomously are particularly promising.
So far proposals and demonstrations of passive interfaces have  dominantly been presented in the framework of cavity QED, such as passive swap operations \cite{Bechler2018PassiveSwapOperation} or passive state transfer \cite{GiannelliSchmit2018} between a photon and an atom.

Recently a general class of resonant dissipation-enabled processes to achieve optimal passive quantum state transfer between the polarization degrees of freedom of a single photon wave packet and a quantum emitter in the adiabatic limit has been proposed theoretically \cite{at15}.  
These processes are capable of turning dominant dissipative processes, such as spontaneous photon emission by a material emitter, into valuable tools for quantum information processing and quantum communication. 
They work for an arbitrarily shaped single-photon
wave packet with sufficiently small bandwidth provided a matching condition is satisfied which balances the
dissipative rates involved. 
In particular, these processes are passive and do not require additional laser pulses or quantum feedback and also do
not require high finesse optical resonators. 
They can be used to enhance significantly the coupling of
a single photon to a single quantum emitter implanted in a one-dimensional waveguide, for example. 
So far, the theoretical discussion of these quantum state transfer processes has concentrated on the quantum dynamics of the emitter which is excited by the one-photon wave packet in the extreme adiabatic limit and decays again by spontaneous emission of this photon. Within this theoretical treatment implementations of a deterministic quantum memory and of a deterministic frequency converter between photonic qubits of different wavelengths have been presented \cite{at15}. Recently, such an adiabatic quantum state transfer process has also received a 'proof-of-concept' demonstration within a cavity-QED scenario \cite{Rempe2020}. 

The main aim of this paper is to extend the previous theoretical work of Ref. \cite{at15} by taking into account and exploiting the properties of the spontaneously emitted photon wave packet which have so far been neglected. Within this generalization also non adiabatic corrections to the extreme adiabatic limit are taken into account systematically. Thereby, we concentrate on scenarios in which the one-photon wave packet before and after the spontaneous emission process is confined to an optical waveguide capable of controlling the propagation directions of the one-photon wave packet by optical circulators. This way it is demonstrated that the photonic degrees of freedoms of these adiabatic one-photon quantum state transfer processes can be used for the passive, heralded and deterministic preparation of Bell states of two material quantum emitters and for realizing a large family of symmetric and asymmetric quantum cloning processes.

This paper is organized as follows.
In Sec.~\ref{sec:Transfer} the basic ideas of the previously introduced  \cite{at15}
dissipation-enabled passive adiabatic quantum state transfer processes between the polarization degrees of freedom of a single photon wave packet and a quantum emitter are summarized. Thereby, these previous investigations are generalized also to weakly non adiabatic cases in which the Fourier-limited bandwidth of the single photon wave packet is no longer negligibly small in comparison with the characteristic dissipative rates involved. The quantum optical model is presented in
Sec.~\ref{sec:TransferSetup}, in Sec.~\ref{sec:TransferCalculation} the resulting quantum state transfer dynamics is discussed, and in Sec. \ref{sec:TransferConditions} the matching conditions of the dissipative rates are introduced which enable optimal quantum state transfer in the extreme adiabatic limit. Basic properties of the quantum state of the spontaneously emitted photon are investigated 
 in Sec.~\ref{sec:TransferEmittedPhoton}.
Theoretical proposals for generating entangled Bell states between two distant material quantum emitters are presented in Sec.~\ref{sec:Entanglement}. Whereas the basic ideas are discussed within a simplified scheme with a linear waveguide in Sec.~\ref{sec:LinearSetup}, a more general scheme for deterministic passive heralded entanglement generation is presented in Sec.~\ref{sec:RingSetup}.
Proposals for implementing symmetric and asymmetric cloning processes capable of cloning an arbitrary pure polarization state of the initially prepared one-photon wave packet onto two material quantum emitters are introduced in Sec.~\ref{sec:Cloning}.
Whereas optimal universal symmetric cloning processes are discussed in
Sec.~\ref{sec:SymmetricCloning}, more general symmetric and asymmetric cloning processes are presented in Sec. \ref{sec:CloningAsymmetric}.

\section{Resonant dissipation-enabled excitation transfer and photon emission} \label{sec:Transfer}

This section summarizes basic properties of the recently presented  adiabatic dissipation-enabled resonant excitation transfer processes \cite{at15} induced by a single-photon wave packet. 
The basic principle of these processes is schematically indicated in Fig.~\ref{fig:schema_transfer}. A quantum emitter initialized in state
 $\ket{s}$ is resonantly and adiabatically excited by a single-photon wave packet
in state $\ket{\Psi_{in}}$ via one of two orthogonal transitions,
i.e. the transition $e \leftrightarrow s$ of Fig.~\ref{fig:schema_transfer}. Adiabatic excitation means that the pulse duration of this photon wave packet is
significantly larger than all other relevant physical time scales.
If the impedance matching condition $\Gamma_{es} = \Gamma_{ef}$ between the two relevant spontaneous decay rates
is fulfilled and if the quantum emitter is fully coupled to the optical waveguide,
the spontaneous photon emission takes place deterministically via the transition $e \leftrightarrow f$
and the quantum emitter ends up in its final state $\ket{f}$. This section extends the investigation of adiabatic dissipation-enabled resonant excitation
transfer processes to higher orders in the adiabatic parameter and discusses basic properties of the spontaneously emitted photon. Whereas implications of this generalization to excitation transfer probabilities are discussed in
Sec.~\ref{sec:TransferConditions}, 
Sec.~\ref{sec:TransferEmittedPhoton} focuses on basic properties of the spontaneously emitted photon.

\subsection{The quantum optical model} \label{sec:TransferSetup}

\begin{figure}[t]
	\includegraphics[scale=1]{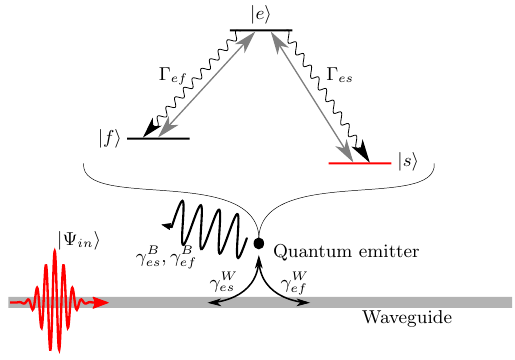}
    \caption{Schematic representation of a quantum emitter coupled to a waveguide: A single photon excites the quantum emitter from its initial state $\ket{s}$ to its excited state $\ket{e}$ from which it can decay spontaneously to the energetically lower lying states $\ket{f}$ and $\ket{s}$ with the spontaneous decay rates $\Gamma_{ef}=\gamma^B_{ef}+\gamma^W_{ef}$ and $\Gamma_{es}=\gamma^B_{es}+\gamma^W_{es}$. The photon can be emitted spontaneously either into the background ($B$) or into the waveguide ($W$) modes.}
    \label{fig:schema_transfer}
\end{figure}

The basic model of dissipation-enabled excitation transfer describes a quantum emitter, modeled by a three-level system, coupled to an electromagnetic continuum of modes. In a free space scenario, for example, this electromagnetic continuum includes all modes of the radiation field coupling to the quantum emitter. 
In a waveguide or optical fiber scenario this electromagnetic continuum predominantly includes the modes inside the waveguide and possibly also additional modes
outside of the waveguide \cite{at15}. With the specific applications presented in the later sections in mind, in this section we concentrate on excitation transfer in a waveguide scenario as depicted schematically in Fig.~\ref{fig:schema_transfer}. 

Let us consider a three-level quantum emitter which is located at position $\textbf{x}_A$ and which is coupled to a single optical photon propagating along a one-dimensional waveguide or optical fiber and prepared initially in state $\ket{\psi_{in}}$ at time $t_0$. Within the dipole- and rotating-wave-approximations the interaction-picture Hamiltonian describing this interacting quantum system is given by 
\begin{align} \label{eq:Hamiltonian}
\!\!\!\!\hat{H}_{int}(t) = &-\left[ e^{-i\omega_{es}(t-t_0)} \textbf{d}_{es}^{\ast}\! \cdot  \hat{\textbf{E}}^-\!(\textbf{x}_A, t) \ket{s}\!\bra{e} + H.c. \right]  \\
    &-\left[ e^{-i\omega_{ef}(t-t_0)} \textbf{d}_{ef}^{\ast} \!\cdot  \hat{\textbf{E}}^-\!(\textbf{x}_A, t) \ket{f}\!\bra{e} + H.c. \right].\nonumber
\end{align}
Thereby $\omega_{es}>0$ and $\omega_{ef}>0$ are the resonance frequencies of the photon-induced transitions $\ket{e} \leftrightarrow \ket{s}$ and $\ket{e} \leftrightarrow \ket{f}$. The corresponding dipole matrix elements characterizing the strengths of these transitions are denoted by  $\textbf{d}_{es} = \bra{e} \hat{\textbf{d}} \ket{s} $ and $\textbf{d}_{ef}  = \bra{e} \hat{\textbf{d}} \ket{f} $. In the interaction picture the negative frequency part of the electric field operator is given by 
\begin{eqnarray} \label{eq:ElectricFieldOperator}
\hat{\textbf{E}}^-(\textbf{x}, t) &=&-i\sum_{\omega,\lambda} \sqrt{\frac{\hbar \omega}{2\epsilon_0}}\textbf{u}^*_{\omega,\lambda}(\textbf{x})a^{\dagger}_{\omega,\lambda}e^{i\omega(t-t_0)}
\end{eqnarray}
with $ \hat{\textbf{E}}^+(\textbf{x}, t) = \left(\hat{\textbf{E}}^-(\textbf{x}, t)\right)^{\dagger}$.
It is determined by the creation operators $a^{\dagger}_{\omega,\lambda}$ of all field modes $(\omega,\lambda)$, each of which is characterized by its frequency and all other quantum numbers, such as direction of propagation and polarization.
The spatial properties of the field modes are described by orthonormal mode functions $\textbf{u}_{\omega,\lambda}(\textbf{x})$ solving the Helmholtz equation with appropriate boundary conditions.
For the explicit modeling of the waveguide setup depicted in Fig.~\ref{fig:schema_transfer} all field modes coupling to the quantum emitter at the position $\textbf{x}_A$ are decomposed into four mode reservoirs, i.e. $W_{es}, W_{ef}, B_{es}, B_{ef}$, whose mode functions are assumed to be orthogonal.
The orthogonal reservoirs $W_{es}$ and $W_{ef}$ contain the modes of interest propagating along the waveguide and centered around the optical transitions frequencies $\omega_{es}$ and $\omega_{ef}$. Analogously, the orthogonal modes $B_{es}$ and $B_{ef}$ are also centered around the optical transitions frequencies
$\omega_{es}$ and $\omega_{ef}$ but
involve all other physically relevant modes. From the
physical applications discussed in the subsequent sections it will become apparent how this partitioning of orthogonal modes is determined by the physical problem of interest in a natural way.
As a result the electric field operator can be split into the contributions of these four orthogonal reservoirs, i.e.
$$ \hat{\textbf{E}}^{\pm}(\textbf{x}, t) = 
\hat{\textbf{E}}^{\pm}_{W_{es}}(\textbf{x}, t) + \hat{\textbf{E}}^{\pm}_{W_{ef}}(\textbf{x}, t) + \hat{\textbf{E}}^{\pm}_{B_{es}}(\textbf{x}, t) + \hat{\textbf{E}}^{\pm}_{B_{ef}}(\textbf{x}, t).$$
It should be mentioned that in general the modes describing the propagation of interest inside a waveguide (here the $W$-modes) are only approximately orthogonal to the field modes describing the background (here the $B$-modes). In general, there are spatial overlaps between these two types of modes due to evanescent waves, for example, with typical exponential tails.  Our assumption of orthogonal waveguide $W$-modes and background $B$-modes applies to all those cases in which such possible overlaps are negligibly small.

Besides waveguide scenarios also free-space of cavity-QED scenarios can be described within this general theoretical framework \cite{at15} by properly identifying the physical significance of the orthogonal modes involved.
In a free-space scenario the reservoir $W_{es}$  contains all the modes of the photon coupling to the transition $\ket{e} \leftrightarrow \ket{s}$, the background reservoir $B_{es}$ contains all other modes coupling to $\ket{e} \leftrightarrow \ket{s}$, the reservoir $W_{ef}$ contains all modes coupling to $\ket{e} \leftrightarrow \ket{f}$ and  the set of modes $B_{ef}$ is chosen to be empty. 
In a cavity-QED scenario a quantum emitter typically couples resonantly to two modes of a high finesse cavity which in turn couples to the electromagnetic continuum of a waveguide. In addition, spontaneous photon emission of the quantum emitter takes place into field modes orthogonal to the cavity and waveguide modes. 
By considering the quantum emitter and the cavity together as a black box the previously described waveguide scenario can also be adapted to this scenario by a proper identification of the physical significance of the field modes involved. 
This way it should also be possible to adapt the results discussed in the subsequent sections to cavity-QED scenarios.

\subsection{Resonant excitation transfer dynamics} \label{sec:TransferCalculation}

The dynamics of a photon propagating inside the waveguide and interacting with the three-level system are governed by the time dependent Schrödinger equation. In the interaction picture it is given by
\begin{equation}
i\hbar \frac{\text{d}}{\text{d}t} \ket{\Psi(t)} \: = \: \hat{H}_{int}(t) \ket{\Psi(t)}.
\end{equation}
Let us solve this Schr\"odinger equation with
the initial condition
\begin{equation}
\ket{\Psi(t_0)} = \ket{s}^{\text{A}}
\ket{\psi_{in}}^{W_{es}} \ket{0}^{B_{es}} \ket{0}^{W_{ef}}\ket{0}^{B_{ef}}
\end{equation}
at time $t_0$.
Hence, at $t_0$ the quantum emitter is initialized in state $\ket{s}^{\text{A}}$ and a single photon state $\ket{\psi_{in}}^{W_{es}} $ is prepared in the waveguide reservoir $W_{es}$ coupling resonantly to the transition $\ket{s}^{\text{A}} \leftrightarrow \ket{e}^{\text{A}}$ (cf. Fig. \ref{fig:schema_transfer}).

Within the rotating wave approximation  the number of excitations is conserved so that the time evolution of the quantum state is of the general form
\begin{eqnarray}\label{eq:Ansatz}
    \ket{\Psi(t)} &= &\Psi_e(t) \ket{e}^{\text{A}}
\ket{0}^{W_{es}} \ket{0}^{B_{es}} \ket{0}^{W_{ef}}\ket{0}^{B_{ef}} \\
&&+ \ket{s}^{\text{A}}
\ket{\Psi_{es}(t)}^{W_{es}, B_{es}} \ket{0}^{W_{ef}}\ket{0}^{B_{ef}} \nonumber\\ 
&&+ 
\ket{f}^{\text{A}}
\ket{0}^{W_{es}} \ket{0}^{B_{es}} \ket{\Psi_{ef}(t)}^{W_{ef},B_{ef}}. \nonumber
\end{eqnarray}
The (unnormalized) single-photon state $\ket{\Psi_{es}(t)}^{W_{es},B_{es}}$ ($\ket{\Psi_{ef}(t)}^{W_{ef},B_{ef}}$) describes the spontaneously emitted photon resulting from the quantum emitter's transition $e\to s$ ($e\to f$) and propagating in the waveguide ($W$) or in the background ($B$) modes. 
The probability of observing the quantum emitter in the excited state $\ket{e}^{\text{A}}$ and the field in the vacuum state of the field modes is given by $|\Psi_e(t)|^2$
with the complex-valued excited-state amplitude $\psi_e(t)$.

The time evolution of this quantum state can be determined with the help of the Weisskopf-Wigner approximation \cite{weisskopf1930wigner}. This approximation is valid as long as all field-induced transition rates of the quantum emitter are small in comparison with the optical transition frequencies involved 
and as long as a spontaneously emitted photon does not interact with the quantum emitter again so that non-Markovian effects are negligible. 
As shown in Appendix \ref{app:DecayEquationDerivation}
the excited-state amplitude $\psi_e(t)$ obeys the differential equation 
\begin{eqnarray} \label{eq:DecayRatePsiE}
    \frac{d}{dt}\Psi_e(t) &=& -\frac{\Gamma}{2}  \Psi_e(t) + i \sqrt{\gamma^W_{es}} f_{in}(t).
\end{eqnarray}
According to the golden rule \cite{FermisGoldenRule} $\Gamma = \gamma^W_{es} + \gamma^B_{es} + \gamma^W_{ef} + \gamma^B_{ef}$ denotes the total decay rate of the quantum emitter state $\ket{e}^{\text{A}}$ due to spontaneous emission of a photon of frequency $\omega_j$ with $j\in\lbrace es, ef\rbrace$ into the possible field reservoirs $R\in\lbrace W,B\rbrace$. 
This total spontaneous decay rate equals the sum of the partial decay rates
\begin{eqnarray}
    \gamma_j^R &=& \frac{2\pi}{\hbar}  
    \sum\limits_{(\omega, \lambda)\in  R_j } \frac{\hbar\omega}{2\epsilon_0}  \delta(\hbar\omega_j-\hbar\omega)\abs{\textbf{d}_j^{\ast}\cdot  \textbf{u}_{\omega_j,\lambda}(\textbf{x}_A) }^2
\end{eqnarray}
into the four orthogonal field reservoirs $R_j\in \{W_{es}, W_{ef}, B_{es}, B_{ef}\}$.
The complex valued one-photon amplitude
\begin{eqnarray} \label{eq:fin}
     \!\!\!\! f_{in}(t) &=& \! \frac{e^{i\omega_{es}(t-t_0)}}{\hbar\sqrt{\gamma^W_{es}}} ~^{W_{es}}\!  \bra{0}\textbf{d}_{es} \cdot  
     \hat{\textbf{E}}_{W_{es}}^{+}(\textbf{x}_A, t)
     \ket{\psi_{in}}^{W_{es}}
\end{eqnarray}
contains all the information about the incoming single photon necessary for determining its interaction with 
the emitter at position $\textbf{x}_A$.
Solving   \eqref{eq:DecayRatePsiE} yields the general solution
\begin{align} \label{eq:PsiEIntegral}
    \Psi_e(t) = i \int_{t_0}^t \sqrt{\gamma^W_{es}} \exp\left( -\frac{\Gamma}{2} (t-t') \right) f_{in}(t') dt'
\end{align}
for the probability amplitude that at time $t$ the quantum emitter is in the excited state $\ket{e}^{\text{A}}$.

\subsection{Optimal adiabatic excitation transfer} \label{sec:TransferConditions}

In the adiabatic limit the emitter's decay rate $\Gamma$ is large in comparison with the effective pulse duration $\tau_{eff}$ of the single photon wave packet approaching the quantum emitter along the waveguide. This effective pulse duration can be measured by the corresponding Fourier-limited effective bandwidth
\begin{equation}
    \Delta\omega = \tau_{eff}^{-1} =  \left( \frac{\int_{-\infty}^{\infty} dt \abs{f'_{in}(t)}^{2}}{\int_{-\infty}^{\infty} dt\abs{f_{in}(t)}^2}   \right)^{1/2}
\end{equation}
so that the adiabatic limit is characterized by the condition $\Delta\omega \ll \Gamma$.
Recent advancements in the area of subnatural-linewidth single photonics \cite{Slattery2019SinglePhotonReview} allow production of single-photon sources below a bandwidth of $1$~MHz, for example. Thus, this adiabatic regime is experimentally accessible if $\Gamma$ is of the order of a typical spontaneous photon emission rate \cite{Liu2020NarrowPhotons}, for example.
As demonstrated in recent experiments this adiabatic regime may be realized by increasing the total spontaneous decay rate $\Gamma$ 
by sufficiently enhancing photon emission into the waveguide \cite{Lund2008EnhancedDecay} or by appropriately tuning the coupling parameters in a cavity-QED scenario \cite{Rempe2020}.

In the adiabatic limit the exponential function in \eqref{eq:PsiEIntegral} decays fast in comparison with the characteristic time scale on which $f_{in}(t')$ changes significantly. Thus, the integral in \eqref{eq:PsiEIntegral}
can be evaluated approximately by iterated partial integration. This yields an asymptotic series  \cite{BenderOrszag} in the small adiabatic parameter $\Delta\omega/\Gamma$. 
In this approximation  the boundary terms at $t_0$ may be neglected as the single photon amplitude at the position of the quantum emitter $x_A$ vanishes initially, i.e. $f_{in}(t_0)=0$, and moreover in the adiabatic limit this amplitude is assumed to rise so slowly that its relevant derivatives of higher order also vanish, i.e. $f_{in}^{(n)}(t_0)=0$, for all $n$ relevant in the asymptotic series.
As a result we arrive at the (formal) asymptotic series
\begin{eqnarray} \label{eq:PsiEFullAnalytic}
    \Psi_e(t)  &=& \frac{i\sqrt{\gamma^W_{es}}}{\Gamma/2}   \sum\limits_{n=0}^{\infty} \left( \frac{-2}{\Gamma}\right)^{n}
    f_{in}^{(n)}(t) 
\end{eqnarray}
for the probability amplitude that the quantum emitter is in its excited state $\ket{e}^{\text{A}}$ at time $t$. 
Together with the field states $\ket{\Psi_{es}(t)}^{W_{es},B_{es}}$ and
$\ket{\Psi_{ef}(t)}^{W_{ef},B_{ef}}$
given by   \eqref{eq:PsiES} and \eqref{eq:PsiEF} of Appendix \ref{app:DecayEquationDerivation} this yields a complete description of the quantum state $\ket{\Psi(t)}$
of  (\ref{eq:Ansatz}) describing the interaction of the single photon with the quantum emitter in the adiabatic limit.
From these expressions 
the probability $p_{s\rightarrow f}$ for successful excitation transfer from state $\ket{s}^{\text{A}}$ to state $\ket{f}^{\text{A}}$ after the interaction  can be determined in a straightforward way, It is given by
\begin{align} \label{eq:TransferProbabilityPsiE}
    p_{s\rightarrow f} =& 
   \Gamma_{ef} \int_{t_0}^{\infty} \abs{\Psi_e(t)}^2 dt \nonumber \\
     =&\frac{4\Gamma_{ef}\gamma^W_{es}}{\Gamma^2}  \int_{t_0}^{\infty} \left| \sum\limits_{n=0}^{\infty} \left( \frac{-2}{\Gamma}\right)^n   f_{in}^{(n)}(t) \right|^2 dt
\end{align}
with 
$\Gamma_{es} = \gamma^W_{es} + \gamma^B_{es}$,~$\Gamma_{ef} = \gamma^W_{ef} + \gamma^B_{ef}$, and $ \Gamma = \Gamma_{es}+\Gamma_{ef}$. 
As demonstrated in Appendix \ref{app:IteratedPartialIntegration},  with the help of successive partial integrations this excitation transfer probability
can be rewritten  in the equivalent form
\begin{align} \label{eq:TransferProbTermsOfFin}
    p_{s\rightarrow f}  =\eta s 
\end{align}
    with
\begin{align}
    s =\sum\limits_{n=0}^{\infty} \left( -\frac{4}{\Gamma^2} \right)^n \int_{t_0}^{\infty}\abs{f_{in}^{(n)}(t)}^2dt \label{eq:TransferProbabilitySumS}
\end{align}
and with the efficiency
\begin{align} \label{eq:Efficiency}
    \eta = \frac{4\Gamma_{ef}\Gamma_{es}}{(\Gamma_{es}+\Gamma_{ef})^2}  \frac{\gamma^W_{es}}{\gamma^W_{es}+\gamma^B_{es}
    }. 
\end{align}
The asymptotic series of \eqref{eq:TransferProbTermsOfFin} generalizes the previous result of  Ref. \cite{at15} by taking into account also all higher order corrections with $n\geq 1$ in the adiabatic limit $\Delta \omega/\Gamma \ll 1$.

The efficiency of \eqref{eq:Efficiency} only depends on the spontaneous decay rates involved and is independent of the pulse form of the initially prepared single-photon wave packet. It becomes maximal if the impedance-matching conditions
\begin{align} \label{eq:ImpedanceMatchingConditions}
    \Gamma_{es} = \Gamma_{ef}, \quad\quad \gamma^B_{es} \ll \gamma^W_{es}
\end{align}
are fulfilled.
The condition $\gamma^B_{es} \ll \gamma^W_{es}$ ensures that the transition $\ket{s}^{\text{A}}\leftrightarrow\ket{e}^{\text{A}}$ predominantly couples to the waveguide reservoir $W_{es}$. Together with the condition $\Gamma_{es}=\Gamma_{ef}$ it balances the decay rates in such a way that the emitted photon interferes with the incoming photon in the $W_{es}$-reservoir in a completely destructive way so that the probability that the photon is emitted spontaneously into the waveguide reservoir $W_{es}$ vanishes $[$cf. also  \eqref{eq:IncidentPhotonNormFiber}$]$. Thus, after the interaction the single photon is in the $W_{ef}$ reservoir  with the quantum emitter in state $\ket{f}^{\text{A}}$ $[$cf. also  \eqref{eq:EmittedPhotonNorm}$]$.
According to \eqref{eq:TransferProbTermsOfFin}, besides the efficiency $\eta$ the probability of excitation transfer $p_{s\rightarrow f}$ after the interaction is also determined by the sum $s$ of  \eqref{eq:TransferProbabilitySumS} involving integrals over the single photon amplitude $f_{in}(t) \equiv f_{in}^{(0)}(t)$ and all its higher order derivatives. Up to second order in the adiabatic parameter $\Delta \omega/\Gamma$, this sum is given by
\begin{align*}
s = \int_{t_0}^{\infty} \abs{f_{in}(t)}^2 dt\left( 1  - \frac{4\Delta\omega^2}{\Gamma^2} + O\left(\left(\frac{4\Delta\omega^2}{\Gamma^2}\right)^2\right) \right).
\end{align*}
If the initially prepared single photon is definitely absorbed, i.e. $\int_{t_0}^{\infty} \abs{f_{in}(t)}^2 dt=1$, in the extreme adiabatic limit
$\Delta \omega \ll \Gamma$
this sum tends to unity and
becomes completely independent of details of the shape of this single-photon amplitude. 
Hence, apart from small corrections of the order of $\Delta\omega^2/\Gamma^2\ll 1$ in this limit perfect excitation transfer, i.e. $p_{s\rightarrow f} \rightarrow 1$, is achievable.
However, an efficient suppression of photon decay into the background is a crucial condition for this purpose.
In a cavity-based scenario a recent experimental implementation \cite{Rempe2020} of adiabatic excitation transfer has been achieved with an excitation probability of 92$\%$.

\subsection{Properties of the emitted photon} \label{sec:TransferEmittedPhoton}

From the field states $\ket{\Psi_{es}(t)}^{W_{es},B_{es}}$ and
$\ket{\Psi_{ef}(t)}^{W_{ef},B_{ef}}$ as
given by   \eqref{eq:PsiES} and \eqref{eq:PsiEF} of Appendix \ref{app:DecayEquationDerivation} and from \eqref{eq:PsiEFullAnalytic} basic properties characterizing the spontaneously emitted photon can be derived. Thus, the probabilities $p^{R_j}$ of detecting the single photon in any of the four reservoirs $R_j\in \{W_{es},W_{ef}, B_{es}, B_{ef}\}$ after the interaction, i.e. at $t\gg 1/\Gamma$, are given by
\begin{equation}
p^{R_j} = \gamma_j^R \int_{t_0}^{\infty} \abs{\Psi_e(t)}^2 dt =  \gamma_j^R~
\frac{p_{s\rightarrow f}}{
   \Gamma_{ef}} \label{eq:EmittedPhotonNorm}
\end{equation}
for $R_j\in \{W_{ef},B_{es}, B_{ef}\}$ 
and by
\begin{align}
p^{W_{es}} 
=&
 1 - \left( \Gamma -\gamma_{es}^W\right)~
\frac{p_{s\rightarrow f}}{
   \Gamma_{ef}} \nonumber \\
=& 1 - \frac{4\gamma^W_{es}}{\Gamma}s + \frac{4(\gamma^W_{es})^2}{\Gamma^2}s . \label{eq:IncidentPhotonNormFiber}
\end{align}

In the following sections we explore situations in which the photon in the reservoir $W_{es}$  has a well defined polarization $\lambda$ and propagates from emitter $\EI$ to a spatially well separated emitter $\EII$ along a linear waveguide. The one-photon amplitude $f_{es}^{\EII}(t)$ characterizing this single-photon wave packet, which eventually excites emitter \EII \:, can be determined analogously to \eqref{eq:fin} with the help of  
\eqref{eq:PsiES}, i.e.
\begin{align}
&f_{es}^{\EII}(t) = \frac{e^{i\omega_{es}(t-t_0)}}{\hbar\sqrt{\gamma^W_{es}}}
	\bra{0}\textbf{d}_{es} \cdot  
     \hat{\textbf{E}}_{W_{es}}^{+}(\textbf{x}_B, t)
     \ket{\Psi_{es}(t)}^{W_{es},B_{es}}
     \nonumber\\
     &=
     e^{i\omega_{es}\L/c}  
       \Big( f^{(0)}_{in}(t- \L/c) + i\sqrt{\gamma^W_{es}}
     \psi_e(t-\L/c) \Big). \label{eq:WesOnePhotonAmplitude}
\end{align}
Thereby, the length of the path between these emitters is denoted by $L_{AB}$, $c$ is the phase velocity of light in the waveguide,
and $\psi_e(t)$ is given by \eqref{eq:PsiEFullAnalytic}.

The distortion experienced by the photon wave packet while propagating in reservoir $W_{es}$ can be characterized by the scalar product of the incident field state $\ket{\Psi_{in}}=\ket{\psi_{in}}^{W_{es}}\ket{0}^{B_{es}}$ with the emitted photon state $\ket{\Psi_{es}(t)}^{W_{es},B_{es}}$ as  given in \eqref{eq:PsiES}. As shown in Appendix \ref{app:ScalarProductRelation} we obtain the result
\begin{align}
    \lim\limits_{t\rightarrow\infty} &\bra{\Psi_{in}} \Psi_{es}(t) \rangle^{W_{es},B_{es}}  = 1 - \frac{\gamma^W_{es}}{\Gamma/2} (s+ir) \label{eq:EmittedIncidentOverlap}
\end{align}
with $s$ defined in \eqref{eq:TransferProbabilitySumS} and with $r$ given by
\begin{align}
    r &= -i \sum\limits_{n=0}^{\infty} \left( \frac{-2}{\Gamma} \right)^{2n+1} \int_{t_0}^{\infty} f_{in}^{\ast}(t)f_{in}^{(2n+1)}dt. \label{eq:SumR}
\end{align}
While $s$ is determined by the adiabatic parameter $\Delta \omega/\Gamma$ the quantity  $r$ captures the odd momenta of the photon's spectral distribution, and vanishes for a time-symmetric pulse, for example. In the extreme adiabatic limit $\Delta \omega/\Gamma \rightarrow 0$ all higher moments vanish, i.e. $r\rightarrow 0$ and $s\rightarrow \int_{t_0}^{\infty}\abs{f_{in}(t)}^2dt$.

Using \eqref{eq:EmittedIncidentOverlap} we can decompose the one-photon amplitude $f_{es}^{\EII}(t)$ of \eqref{eq:WesOnePhotonAmplitude} into the amplitude $f_{in}^{(0)}(t)$ of the initial photon wave packet $\ket{\Psi_{in}}$ and an amplitude $f_\perp(t)$ corresponding to an orthogonal component, i.e.
\begin{align}
    f_{es}^{\EII}(t) =& \frac{\gamma^W_{es}}{\Gamma/2}\sqrt{s- s^2-r^2} f_\perp(t) \label{eq:WesOrthogonalDecomposition} \\
    +& \left( 1 - \frac{\gamma^W_{es}}{\Gamma/2} (s+ir) \right) e^{i\omega_{es}\L/c}f^{(0)}_{in}(t- \L/c).\nonumber 
\end{align}
The orthogonal contribution $f_\perp(t)$ is fully determined by $f_{es}^{\EII}(t)$ and by $f_{in}^{(0)}(t)$ and depends on the coupling of the emitters to the specific mode structure under consideration. It is apparent that in the extreme adiabatic limit, i.e. $\Delta \omega/\Gamma \rightarrow 0$, the contribution of this orthogonal amplitude vanishes provided $s \rightarrow \int_{t_0}^{\infty}\abs{f_{in}(t)}^2dt=1$.
Under these conditions
\eqref{eq:WesOrthogonalDecomposition} reduces to the result
\begin{align}
f_{es}^{\EII}(t) = \sqrt{p^{W_{es}}} e^{i\omega_{es}\L/c}f^{(0)}_{in}(t- \L/c). \label{eq:PhotonAmplitudeAdiabatic}
\end{align}
Thus, the orthogonal component becomes vanishingly small so that under these conditions the spontaneously emitted photon is not distorted with respect to the incoming photon. 
It is apparent that \eqref{eq:WesOnePhotonAmplitude} and \eqref{eq:WesOrthogonalDecomposition} are consistent with the conservation of probability, i.e.
\begin{eqnarray}
\int_{t_0}^{\infty} \vert f_{es}^{\EII}(t)\vert^2dt &=&
p^{W_{es}}.
\end{eqnarray}

We may also consider a scenario in which the photon is spontaneously emitted into the reservoir $W_{ef}$ and propagates along a linear waveguide to position $\textbf{x}_D$. Analogous to  \eqref{eq:WesOnePhotonAmplitude}, in this case the one-photon amplitude $f_{ef}^{D}(t)$ characterizing the single-photon wave packet $\ket{\Psi_{ef}(t)}^{W_{ef},B_{ef}}$ of \eqref{eq:PsiEF} reads
\begin{align}
f_{ef}^{D}(t) =& i\sqrt{\gamma^W_{ef}}e^{i\omega_{ef}\LL/c}
     \psi_e(t-\LL/c).
     \label{eq:WefOnePhotonAmplitude}
\end{align}
In the extreme adiabatic limit \eqref{eq:WefOnePhotonAmplitude} simplifies to
\begin{align}
f_{ef}^{D}(t) =-\sqrt{p^{W_{ef}}}~e^{i\omega_{ef}\LL/c}f^{(0)}_{in}(t- \LL/c). \label{eq:WefAmplitudeAdiabatic}
\end{align}
Thus, in the extreme adiabatic limit the one-photon amplitude \eqref{eq:WefAmplitudeAdiabatic} of the photon emitted into the $W_{ef}$ reservoir after successful excitation transfer differs from the amplitude \eqref{eq:PhotonAmplitudeAdiabatic} by a phase shift of $\pi$.

\section{Entanglement Generation} 
\label{sec:Entanglement}

As discussed in the previous section, the pulse shape of a single-photon wave packet is not changed under dissipation-enabled excitation transfer  in the extreme adiabatic limit. This property can be exploited for entanglement generation between two distant quantum emitters as their emitted photons become indistinguishable. In this section
we present first a basic probabilistic scheme for entanglement generation between two distant quantum emitters positioned along a linear waveguide in Sec.~\ref{sec:LinearSetup}. Then in Sec.~\ref{sec:RingSetup}
a more general deterministic scheme is presented for preparing Bell states in a waveguide ring with the help of optical circulators.

\subsection{Entanglement generation along a linear waveguide} \label{sec:LinearSetup}

\begin{figure}[t]
	\includegraphics[scale=1]{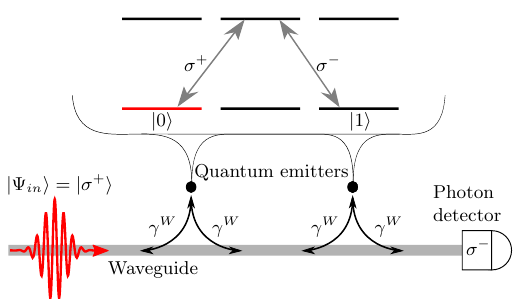}
    \caption{Schematic representation of entanglement generation: A $\sigma^+$-polarized single-photon wave packet propagates (from left to right) along a linear waveguide and interacts with two identical quantum emitters, thereby causing adiabatic excitation transfer. After successful excitation transfers at both emitters detection of a $\sigma^-$-photon at the detector heralds entanglement between the quantum emitters as it is impossible to discriminate which emitter has emitted a $\sigma^-$-polarized photon spontaneously. The level schemes of the quantum emitters are indicated with initial states in red and final states in black. Due to dipole selection rules the central spontaneous decay channel is forbidden for photons propagating along the linear waveguide.}
    \label{fig:schema_entanglement_conceptual}
\end{figure}

We consider an excitation scenario as depicted schematically in Fig.~\ref{fig:schema_entanglement_conceptual}.
Two identical quantum emitters, say $\EI$ and $\EII$, can perform optical transitions between two threefold degenerate manifolds of unit angular momenta. These emitters are assumed to be perfectly coupled to electromagnetic field modes inside a linear waveguide with a well defined propagation direction.  This direction will be used as the quantization axis in our subsequent discussion. Initially a  circularly polarized single-photon wave packet, say with $\sigma^+$ polarization, is prepared inside this waveguide and propagates towards the quantum emitters. 
These emitters are assumed to be spatially separated by a distance $\L$ large in comparison with the spatial extension of the initially prepared single-photon wave packet, i.e. $\L = \vert \textbf{x}_{\EI} - \textbf{x}_{\EII}\vert \gg c/\Delta \omega$. Thereby, $c$ denotes the phase velocity inside the linear waveguide and $\Delta \omega$ is the effective bandwidth of the wave packet. Therefore, both quantum emitters can be excited adiabatically in such a way that the resulting excitation transfer processes are well separated in time. 
For instance, a single photon bandwidth of $\Delta\omega / (2\pi) = 5~{\rm s}^{-1}$ requires a minimum distance between the emitters of at least $60$~m. A proof-of-principle experiment demonstrating the feasibility of such adiabatic dissipation-enabled resonant processes over this distance was performed recently by Daiss et al. \cite{Rempe2021CNOTEntanglement}, for example.

Initially both emitters are 
prepared in their  initial states $\ket{0}^{\EI}$ and $\ket{0}^\EII$. Each of these initial states is part of the ground state manifold and couples to the excited state manifold by the initially prepared circularly polarized $\sigma^+$-modes. In each emitter one of the other ground states, i.e. $\ket{1}^\EI$ and $\ket{1}^\EII$, couples to the same excited state by the orthogonal circularly polarized $\sigma^-$-modes. 
Provided the coupling of the emitters to the electromagnetic field modes of the linear waveguide is perfect, spontaneous photon emission from the excited emitter state
to the ground state manifold cannot involve a photon linearly polarized along the propagation direction of the wave packet, i.e. the quantization axis. Therefore, spontaneous photon emission from the excited central emitter state to the central ground state is forbidden by dipole selection rules so that the level scheme realizes an ideal lambda system as discussed in Sec.~\ref{sec:Transfer}. 

In the following it is convenient to decompose the radiation field inside the linear waveguide into four reservoirs, i.e. $W_{\leftarrow}^\pm$ and $W_{\rightarrow}^\pm$.
These reservoirs contain the modes describing photon propagation to the left ($\leftarrow$) or to the right  ($\rightarrow$) with
$\sigma^+$ ($+$) or $\sigma^-$ ($-$) polarization.
This decomposition ensures that all reservoirs  are pairwise orthogonal so that we can describe this scenario by the theoretical framework developed in Sec.~\ref{sec:Transfer} with the help of the identifications
$W_{es} = W_{\rightarrow}^{+}$,
$W_{ef} = W_{\rightarrow}^{-}$,
 $B_{es} = W_\leftarrow^+$ and $ B_{ef} = W_\leftarrow^-$. 
 Accordingly, the modes describing propagation to the right, i.e. to the photon detector of  Fig.
\ref{fig:schema_entanglement_conceptual}, constitute the 'waveguide' (W) modes and the modes describing propagation to the left constitute the 'background' (B) modes. As we are neglecting spontaneous photon emission out of the waveguide these reservoirs yield a complete description of the relevant electromagnetic field. Effects of such an emission are discussed at the end of the section. It should be pointed out that this structure of the relevant reservoirs implies that $\int_{t_0}^{\infty} \vert f_{in}(t)\vert^2 dt=1$ so that in the extreme adiabatic limit, i.e. $\Delta \omega/\Gamma \to 0$, the initially prepared single photon is definitely absorbed after the interaction with the first quantum emitter $\EI$. 

Dipole selection rules imply that the 
dipole transitions of the identical emitters couple equally strongly to their respective reservoirs so that the corresponding spontaneous decay rates are equal, i.e. $\gamma_{es}^W = \gamma_{ef}^W= \gamma_{es}^B = \gamma_{ef}^B = \gamma^W$. 
Hence the total decay rate of the excited emitter state is given by $\Gamma = 4\gamma^W$.  
As a result the efficiency for excitation transfer of quantum emitter $\EI$ is given by
$[$cf. \eqref{eq:Efficiency}$]$ 
\begin{align} \label{eq:EntangelementEfficiency}
\eta = 
\frac{\gamma^W}{ 2\gamma^W } = \frac{1}{2}  .
\end{align}
This is due to the fact that the excited emitter state can decay spontaneously also by emission of a $\sigma^+$ photon. Nevertheless, in the extreme adiabatic limit the initially prepared single photon wave packet excites the quantum emitter $\EI$ with unit probability so that in this limit the transitions probability for
excitation transfer is given by $p_{s\to f} = \eta = 1/2$ $[$cf. \eqref{eq:TransferProbTermsOfFin}$]$.

In order to entangle the two quantum emitters we consider two subsequent adiabatic excitation processes. 
Initially both quantum emitters are prepared in their ground states $\ket{0}^\EI$  and $\ket{0}^\EII$. A single  photon wave packet,
initially prepared in the $W_{es} = W_{\rightarrow}^+$ reservoir in the quantum state $\ket{\sigma^+}^\rightarrow$, propagates
to the right towards the first emitter $\EI$. Correspondingly, at time $t_0$ the initially prepared pure quantum state of the two quantum emitters and of the radiation field is given by
\begin{align}
    \ket{\Psi(t_0)} = \ket{\sigma^+}^\rightarrow \ket{0}^{\EI}\ket{0}^{\EII}. \label{eq:InitialEntanglement}
\end{align} 
As the single photon propagates along the waveguide it can cause excitation transfers at both quantum emitters  with probabilities $p_{s\rightarrow f} =\eta s$ $[$cf. \eqref{eq:TransferProbTermsOfFin}$]$ with $s$ being determined by the one-photon amplitude at the location of the corresponding emitter $[$cf. \eqref{eq:TransferProbabilitySumS}$]$. Each excitation transfer flips the polarization of the photon. Therefore,
 detecting a photon within the waveguide behind the second emitter $\EII$ with flipped polarization heralds the occurrence of one of two possible excitation transfer processes either at quantum emitter $\EI$ or at quantum emitter $\EII$. As these two events cannot be distinguished by this photon detection process both quantum emitters become entangled conditioned on the detection of a $\sigma^-$ photon in the reservoir $W_{ef}$.

Let us explore this basic idea of entanglement generation quantitatively.
After interaction with the first emitter $\EI$ the photon can remain in the initial reservoir $W_{es}=W^+_\rightarrow$ with probability $p^{W_{es}}=1-2\eta s + \eta^2s$ $[$cf. \eqref{eq:IncidentPhotonNormFiber}$]$. Its normalized quantum state $\ket{\sigma^+_{\EI}(t)}^{\rightarrow}$ is obtained by projecting the state of \eqref{eq:PsiES} onto the reservoir $W_{es}$. Alternatively,
the photon is spontaneously emitted into any of the other three reservoirs with probabilities $p^{W_{ef}} = p^{B_{es}}=p^{B_{ef}} = \eta^2 s$ $[$cf. \eqref{eq:EmittedPhotonNorm}$]$. 
In case of successful excitation transfer the resulting $\sigma^-$-polarized photon state in reservoir $W_{ef}$ is denoted by $\ket{\sigma^-_\EI(t)}^{\rightarrow}$ and can be determined from \eqref{eq:PsiEF}. Furthermore, let us describe the normalized quantum state of the emitter-field system resulting from
cases in which the photon ends up in the background modes $B_{es}$ or $B_{ef}$ propagating away from the second emitter $\EII$ by  $\ket{\Psi_\text{back}(t)}^{\leftarrow}$. 
As a result
long after the first interaction, i.e. at time $t$ with $(t-t_0)\gg 1/\Gamma$,
the quantum state of the emitter-field system $\ket{\Psi(t)}$
is a linear superposition of three orthogonal quantum states, i.e.
\begin{align}
    \ket{\Psi(t)} &=\sqrt{p^{W_{ef}}} \ket{\sigma^-_\EI(t)}^\rightarrow \ket{1}^\EI\ket{0}^\EII \nonumber\\
    &+ \sqrt{p^{W_{es}}} \ket{\sigma^+_{\EI}(t)}^\rightarrow \ket{0}^\EI\ket{0}^\EII \nonumber\\
    &+ \sqrt{p^{B_{es}}+p^{B_{ef}}} \ket{\Psi_\text{back}(t)}^{\leftarrow}.
\end{align}
It should be mentioned that due to dipole selection rules
the spontaneously emitted photon in quantum state $\ket{\sigma^-_\EI(t)}^\rightarrow$ 
does not interact with the second emitter $\EII$ initialized in state $\ket{0}^{\EII}$. In this case the photon propagates directly to the right to the photon detector located at position  $\textbf{x}_D$. Its one-photon amplitude at the position of the photon detector is given by \eqref{eq:WefOnePhotonAmplitude}.

For interaction with the second emitter $\EII$ only the $\sigma^+$-polarized component $\ket{\sigma^+_{\EI}(t)}^\rightarrow$ in the $W_{es}=W_\rightarrow^+$ reservoir has to be considered. 
According to \eqref{eq:WesOrthogonalDecomposition} at the position of emitter $\EII$ its one-photon amplitude $f_{in}^\EII(t)$
can be separated into a part identical to the initial amplitude $f_{in}(t-L_{AB}/c)$ and an amplitude $f_\perp(t)$ corresponding to an orthogonal component, i.e.
\begin{align}
    f_{in}^{\EII}(t) =&e^{i\omega\L/c}\Big( \left( 1 - \eta (s+ir) \right) f_{in}(t- \L/c) \nonumber\\
    &\quad+ \eta\sqrt{s- s^2-r^2} f_\perp(t)\Big).
\end{align}
Accordingly, the interaction with emitter $\EII$ originates from a superposition of two one-photon amplitudes.
However, in the extreme adiabatic limit, i.e. $s\rightarrow 1$ and $r\rightarrow 0$, the contribution $f_\perp(t)$ of the orthogonal component vanishes.
Since we assume identical couplings of both emitters to the reservoirs the excitation transfer induced by $f_{in}(t - \L/c)$ can be treated similarly as the excitation transfer at the first emitter $\EI$.  Analogously, we denote the normalized quantum state of the  spontaneously emitted $\sigma^-$-polarized photon of emitter $\EII$ by $\ket{\sigma^-_\EII(t)}^\rightarrow$. Its one-photon amplitude at the photon detector is given by
\begin{eqnarray}
f_{ef}^{D}(t) &=& i\sqrt{\gamma^W}e^{i\omega(\L+L_{\EII D})/c}
\psi_e(t-(\L+L_{\EII D})/c).\nonumber\\
\end{eqnarray}
It is identical to the one-photon amplitude of the spontaneously emitted photon of emitter $\EI$ at the position of the photon detector which
originates from the photonic quantum state $\ket{\sigma^-_\EI(t)}^\rightarrow$, as derived in \eqref{eq:WefOnePhotonAmplitude}.

Let us now consider the small contribution of the
amplitude $f_{\perp}(t)$ which vanishes in the extreme adiabatic limit. This amplitude induces an excitation transfer in emitter $\EII$ with probability $\eta s_{\perp}$, where $s_\perp \in \left[ 0,1\right]$ encodes the amplitude analogously to \eqref{eq:TransferProbabilitySumS} but now with the one-photon amplitude $f_\perp(t)$.  This excitation transfer causes spontaneous emission of a photon
with probability $p^{W_{ef}}_\perp = \eta^2s_\perp$. The resulting
normalized single-photon quantum state  $\ket{\sigma^-_{\perp}(t)}^\rightarrow$ is orthogonal to the single-photon quantum state
$\ket{\sigma^-_\EII(t)}^\rightarrow$
since excitation transfer is a unitary process and preserves orthogonality.

As a result, long after both interactions the quantum state of the emitter-field system has the general form
\begin{align}
    \ket{\Psi_{\text{fin}}(t)} &= \sqrt{p^{W_{ef}}} \ket{\sigma^-_\EI(t)}^\rightarrow \ket{1}^\EI\ket{0}^\EII \nonumber\\ 
    &+\sqrt{p^{W_{ef}}} \left(1-\eta (s+ir)\right) \ket{\sigma^-_\EII(t)}^\rightarrow \ket{0}^\EI\ket{1}^\EII\nonumber\\
    &+ \sqrt{p^{W_{ef}}_\perp \eta^2 (s - s^2 - r^2)} \ket{\sigma^-_{\perp}(t)}^\rightarrow \ket{0}^\EI\ket{1}^\EII\nonumber\\
    &+\sqrt{1-R}\ket{\Psi_\text{other}(t)}^{\rightarrow,\leftarrow} .
\end{align}
Thereby, all contributions not involving the reservoir $W_{ef}=W_\rightarrow^-$ are described by the normalized quantum state
$\ket{\Psi_\text{other}(t)}^{\rightarrow,\leftarrow}$. These latter contributions do not cause detection of a
$\sigma^-$-polarized photon. 

In order to determine the quantum state of the two quantum emitters conditioned on the detection of a $\sigma^-$-polarized photon we employ Glauber's one-atom photodetection model \cite{glauber1963detection}.
As the one-photon amplitudes originating from the quantum states 
$\ket{\sigma^-_\EI(t)}^\rightarrow$,
$\ket{\sigma^-_\EII(t)}^\rightarrow$ and 
$\ket{\sigma^-_{\perp}(t)}^\rightarrow$
are not distinguished by the
photodetector positioned at $\textbf{x}_D$, 
detection of a $\sigma^-$-polarized photon after both interactions implies preparation of both quantum emitters in the pure entangled quantum state
\begin{eqnarray}
 \ket{\Psi_{ent}}&=&\left(
 \sqrt{p^{W_{ef}}}\left( \ket{1}^\EI\ket{0}^\EII +  (1-\eta(s+ir)) \ket{0}^\EI\ket{1}^\EII \right)\right.\nonumber\\
 &+&\left.
  \sqrt{p^{W_{ef}}_\perp \eta^2 (s - s^2 - r^2)} \ket{0}^\EI\ket{1}^\EII \right)/\sqrt{R}
\end{eqnarray}
with
probability 
\begin{equation}
  R = \eta^2 s 
  \left(1+(1-\eta s)^2+\eta^2r^2\right)
  + \eta^4 s_\perp (s-s^2-r^2).
\end{equation}
Thus, in the extreme adiabatic limit, i.e. $s\to 1, r\to 0$,  neglecting  terms of the order of $(\Delta \omega/\Gamma)^2$
the quantum state of both emitters conditioned on the postselection of a $\sigma^-$-polarized photon becomes
\begin{align}
\ket{\Psi_{ent}} &=\left(
 \eta \ket{1}^\EI\ket{0}^\EII +  \eta(1-\eta) \ket{0}^\EI\ket{1}^\EII\right)/\sqrt{R}
\end{align}
with the photon detection probability $R \to \eta^2 (1 + (1-\eta)^2)$.  As our setup employs an efficiency of $\eta = 1/2$, in this limit the photon detection probability approaches the value $R = 5/16$.

In order to achieve perfect entanglement in the form of a Bell state, for example, a higher efficiency $\eta$ would be required at the second emitter $\EII$ than at the first emitter $\EI$. Furthermore, it would be desirable to increase the photon detection probability $R$ in order to achieve a higher probability for the conditional preparation of an entangled quantum emitter state. In the following it is demonstrated that both desiderata can be achieved by  interconnecting the emitters by a ring shaped waveguide, whose photon propagation characteristics can locally still be well approximated by a linear waveguide, and by controlling the directions of photon propagation with the help of optical circulators. By this setup it is even possible to achieve perfect entanglement in a deterministic way.

\begin{figure}[t]
	\includegraphics[scale=1]{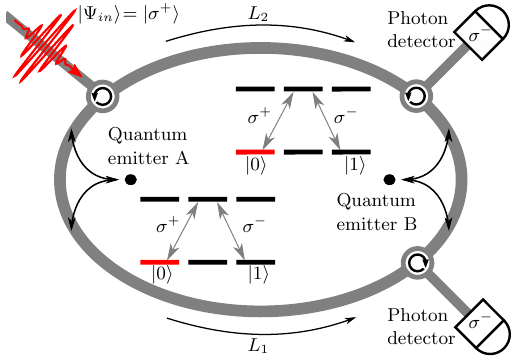}
	\caption{Schematic representation of deterministic entanglement generation in a ring-shaped waveguide: A $\sigma^+$-polarized single-photon wave packet propagates 
    along optical paths indicated by the optical circulators $({\protect\rotatebox[origin=c]{90}{\footnotesize $\circlearrowleft$}}, {\protect\rotatebox[origin=c]{270}{\footnotesize $\circlearrowright$}})$. 
    After successful excitation transfers at both emitters $\EI$ and $\EII$, detection of a $\sigma^-$-polarized photon at one of the two detectors heralds the preparation of a Bell state of the two quantum emitters as each detector cannot distinguish  which emitter has emitted the $\sigma^-$-polarized photon. The level schemes of the quantum emitters are indicated with initial states in red and final states in black. Due to dipole selection rules the central spontaneous decay channel is forbidden.
    }
    \label{fig:schema_entanglement}
\end{figure}

\subsection{Deterministic entanglement generation in a ring-shaped setup}\label{sec:RingSetup}

In this section
 a ring-shaped setup is discussed which enables one-shot Bell state generation by balancing the spontaneous decay rates appropriately and by controlling the photon propagation with the help of optical circulators.
We consider a waveguide ring as schematically shown in 
Fig.~\ref{fig:schema_entanglement}. It
interconnects two identical quantum emitters by two paths of lengths $L_1$ and $L_2$ and involves three optical circulators which send an incoming photon only into the direction indicated by the circular arrows.
The basic excitation and detection mechanisms of this ring-shaped setup  are analogous to the linear model discussed in the previous section. Again it is assumed that the emitters are well separated spatially. In addition it is assumed that the curvature of this long waveguide is so small that locally wave propagation inside this waveguide can be approximated well by wave propagation inside a linear waveguide.

Modifying the notation of the previous section accordingly
the waveguide reservoirs are now called $W^\pm_{\fc}$ and $W^\pm_{\fa}$ and denote the clockwise  ($\fc$) and anticlockwise ($\fa$) propagating modes with $\sigma^+$ or $\sigma^-$ polarization.
Initially the emitter-field system is prepared in the pure quantum state
\begin{eqnarray}
 \ket{\Psi(t_0)} &=& \ket{\sigma^+}^{\fa} \ket{0}^{\EI}\ket{0}^{\EII}.
\end{eqnarray}
The initially prepared $\sigma^+$-polarized single-photon wave packet is injected via an optical circulator so that it propagates in anticlockwise direction, i.e. inside the $W^+_{\fa}$ reservoir.
In order to describe the interaction with the first emitter $\EI$ in the the framework of Sec.~\ref{sec:Transfer} we identify the reservoirs $W_{es}$ and $ W_{\fa}^+$. The background reservoir $B_{es}$ of the transition between the initial state $|0\rangle^\EI$  and the excited state is identified with
the reservoir $W_{\fc}^+$. Furthermore, as both propagation directions of the spontaneously emitted photon lead to the second emitter $\EII$, we choose the target reservoir as $W_{ef} = W_{\fa}^- \cup W_{\fc}^-$ and the corresponding background to be empty, i.e. $B_{ef} = \emptyset$. Again we neglect photon emission out of the ring-shaped waveguide and refer for a discussion of these effects to the end of the section.
With this choice of reservoirs the relevant spontaneous decay rates fulfill the relations $\gamma^W_{es} = \gamma^B_{es} = \gamma^W$, $\gamma^W_{ef} = 2\gamma^W$ and $\gamma^B_{ef}=0$ so that the resulting total decay rate of the excited state is given by
$\Gamma = 4\gamma^W$. The resulting efficiency of adiabatic excitation transfer is given by $\eta = \gamma^W/2\gamma^W = 1/2$. In the extreme adiabatic limit the corresponding transfer probability is given by $p_{s\to f}=1/2$. 
The probability for the spontaneously emitted photon to remain in the reservoir $W_{es}=W_{\fa}^+$ after this excitation transfer is given by $p^{W_{es}}= 1-2\eta s + \eta^2s$ $[$cf. \eqref{eq:IncidentPhotonNormFiber}$]$. According to  \eqref{eq:EmittedPhotonNorm} the probability for spontaneous photon emission into any of the other waveguide reservoirs is given by $p^{W_{ef}}=2p^W$ and $p^{B_{es}}=p^W$
with $p^W = \eta^2s$.

The normalized field states of a spontaneously emitted photon $\ket{\sigma^-_\EI(t)}^{\fc}$ and  $\ket{\sigma^-_\EI(t)}^{\fa}$ can be determined from \eqref{eq:PsiEF} by projection onto the corresponding reservoirs. Analogously the normalized field states $\ket{\sigma^+_\EI(t)}^{\fa}$ and  $\ket{\sigma^+_\EI(t)}^{\fc}$ can be obtained from \eqref{eq:PsiES}. 
Therefore,  
at times $(t-t_0)\gg 1/\Gamma$ long after the interaction with emitter $\EI$ but before interaction with emitter $\EII$ 
the pure quantum state of the emitter-field system
is given by
\begin{align}
    \!\!\!\ket{\Psi(t)} &= \sqrt{p^{W}} \Big(\ket{\sigma^-_\EI(t)}^{\fa}+\ket{\sigma^-_\EI(t)}^{\fc} \Big) \ket{1}^\EI\ket{0}^\EII \nonumber \\ 
    +& \Big( \sqrt{p^{W_{es}}} \ket{\sigma^+_r(t)}^{\fa}+\sqrt{p^W}\ket{\sigma^+_r(t)}^{\fc}\Big)\ket{0}^\EI\ket{0}^\EII \!\!.
\end{align}
The wave packet components of the spontaneously emitted photon
propagate in both directions inside the ring-shaped waveguide along the arms of lengths $L_1$ ($\fa$) and $L_2$ ($\fc$) towards quantum emitter $\EII$. 
In order to describe the interaction of the spontaneously emitted photon with emitter $\EII$ within the theoretical framework of
Sec. \ref{sec:Transfer} we identify the relevant mode reservoirs in the following way: $W_{es} = W_{\fa}^+ \cup W_{\fc}^+$, $W_{ef} = W_{\fa}^- \cup W_{\fc}^-$ and $B_{es} = B_{ef} = \emptyset$.
The corresponding spontaneous decay rates fulfill the relations $\gamma^W_{es} = \gamma^W_{ef} = 2\gamma^W$ and  $\gamma^B_{es} = \gamma^B_{ef} = 0$.
Thus, according to \eqref{eq:Efficiency} the efficiency for excitation transfer at emitter $\EII$ is given by $\eta' = 2\eta = 1$. This is due to the fact that the $\sigma^+$-polarized  components of the spontaneously emitted photon of emitter $\EI$ interfere constructively at the position of the quantum emitter $\EII$. 

Analogously to 
\eqref{eq:WesOnePhotonAmplitude} and \eqref{eq:WefOnePhotonAmplitude}, at the position of emitter \EII \: long after interaction with emitter $\EI$ the one-photon amplitude of the field modes in the $W_{es}$ reservoir 
is given by
\begin{align}
    f_{in}^\EII(t) =& \frac{e^{i\omega L_1/c}}{\sqrt{2}} \Big(f_{in}(t- L_1 /c) + i\sqrt{\gamma^W}
     \psi_e(t-L_1/c) \Big)\nonumber\\
    &\hspace{40pt}+ \frac{e^{i\omega L_2/c}}{\sqrt{2}}  i\sqrt{\gamma^W}
     \psi_e(t-L_2/c). 
\end{align}
If the interfering fractions of the photon wave packet arrive at emitter $\EII$ simultaneously, i.e. $\vert L_1 - L_2 \vert \ll c/\Delta\omega$, 
and the emitter positions $\textbf{x}_{\EI}$ and 
$\textbf{x}_{\EII}$ fulfill the condition of complete destructive interference, i.e.
$e^{i\omega(L_1-L_2)/c} = -1$,
this one-photon amplitude becomes  proportional to the initially prepared time-shifted one-photon amplitude, i.e.
\begin{eqnarray}
f_{in}^\EII(t) &
    =& \frac{e^{i\omega L_1/c}}{\sqrt{2}}f_{in}(t- L_1 /c).\label{eq:finRing}
\end{eqnarray}
In view of recent experimental advancements \cite{Pregnolato2020QuantumDotLocalization, Hood2016InterferenceCondition} the
necessary control over the path lengths $L_1$ and $L_2$ of the ring-shaped waveguide ensuring complete destructive interference is feasible.
In contrast to the setup along a line considered in Sec.~\ref{sec:LinearSetup}, the orthogonal distortions of both pulse components cancel themselves out at the location of emitter $\EII$.
The relation
\begin{eqnarray}
\int_{t_0}^{\infty} \vert f_{in}^{\EII}(t)\vert^2dt &=&
\frac{1}{2}
\end{eqnarray}
implies that
in the extreme adiabatic limit
the transfer probability of emitter $\EII$ is 
given by $p_{s\to f} = \eta' \times 1/2 = 1/2$. 
The probability for photon emission into the waveguide reservoirs $W_{\fa}^-$ and $W_{\fc}^-$ is given by $p^{W_{ef}}/2 = \eta^2s = p^W $ $[$cf. \eqref{eq:EmittedPhotonNorm}$]$.

As a result after the second interaction with emitter $\EII$ the pure quantum state of the emitter-field system is of the form
\begin{align}
    \ket{\Psi_\text{fin}(t)} &= \sqrt{p^{W}}\big( \ket{\sigma^-_\EI(t)}^{\fa}+\ket{\sigma^-_\EI(t)}^{\fc}\big) \ket{1}^\EI\ket{0}^\EII \nonumber \\
    &+\sqrt{p^{W}}\big( \ket{\sigma^-_\EII(t)}^{\fa}+\ket{\sigma^-_\EII(t)}^{\fc}\big) \ket{0}^\EI\ket{1}^\EII \nonumber  \\
    &+  \sqrt{1-R}\ket{\Psi_\text{other}(t)}^{\fa, \fc} 
    \label{eq:EntanglementRingFinal}
\end{align}
with the normalized state $\ket{\Psi_\text{other}(t)}^{\fa, \fc}$ describing all events with unsuccessful excitation transfer. 

The optical circulators ensure that after interaction with emitter $\EI$ or $\EII$ the spontaneously emitted $\sigma^-$-polarized photon can be detected by one of the two photon detectors (cf. Fig.~\ref{fig:schema_entanglement}).
By arguments analogous to the ones presented in Sec.~\ref{sec:LinearSetup} the photon states resulting from emitter $\EII$ with $\sigma^-$ polarization are indistinguishable from the corresponding ones resulting from emitter $\EI$.
Thus, at the position of the ${\fa}$-detector the one-photon amplitudes of the photon states $\ket{\sigma^-_\EI(t)}^{\fa}$ and $\ket{\sigma^-_\EII(t)}^{\fa}$ are equal as the latter involves a phase factor $e^{i\omega L_1/c}$ $[$cf. \ref{eq:finRing}$]$ and the former involves additional propagation along the distance $L_1$ separating both emitters.
At the position of the ${\fc}$-detector the one-photon amplitudes of the photon states
 $\ket{\sigma^-_\EI(t)}^{\fc}$ and $\ket{\sigma^-_\EII(t)}^{\fc}$ differ by a phase of magnitude $\pi$ as the latter involves a phase
 factor $e^{i\omega L_1/c}$ and
 the former involves additional propagation along the distance $L_2$ separating both emitters.
As a result conditioned on the detection of a $\sigma^-$-polarized photon the two emitters are prepared in
one of the Bell states
\begin{align}
    \ket{\Psi_{ent}^{\fa}} &=  \big( \ket{0,1} + \ket{1,0} \big) /\sqrt{2}, \label{eq:PsiEntFa} \\
    \ket{\Psi_{ent}^{\fc}} &=  \big( \ket{0,1} - \ket{1,0} \big) /\sqrt{2} \label{eq:PsiEntFc}
\end{align}
depending on whether the $\fa$-detector or the $\fc$-detector has detected the $\sigma^-$-polarized photon. 
The detection probability is equal for both detectors and is given by
\begin{align}
    R = 4p^W = 4\eta^2 s \label{eq:EntanglementRate}.
\end{align}
Since our setup yields $\eta=1/2$, in the extreme adiabatic limit, i.e. $s\rightarrow 1$,  this detection probability approaches unity. 
Thus, under extreme adiabatic conditions one of these Bell states can be prepared deterministically by this setup. 

In order to demonstrate the possibility of deterministic Bell state generation in this ring-shaped setup we have assumed that the quantum emitters couple to the waveguide perfectly and that photon emission outside of the waveguide is negligible. Let us comment finally on possible effects of photon emission outside of the waveguide within a simplified model. For this purpose let us assume that the excited emitter states can decay spontaneously with a decay rate $\gamma^\text{other}$
also to an energy level outside of the state manifold considered so far. Thus, it is assumed that the resulting spontaneously emitted photon has a different frequency so that its field modes are orthogonal to the field modes already taken into account so far.
In this case adiabatic excitation transfer can still be described in analogy to
Sec.~\ref{sec:Transfer} with the replacement $\Gamma \mapsto \Gamma + \gamma^\text{other}$ thus yielding a modified efficiency $\tilde{\eta}$ for adiabatic excitation transfer, i.e.
\begin{align}
    \eta \mapsto \bar{\eta} =  \frac{2\gamma^W}{4\gamma^W+\gamma^\text{other}}. \label{eq:EfficiencyError}
\end{align}
Thus, the ideal success probability for entanglement generation as described by \eqref{eq:EntanglementRate} decreases accordingly but the conditionally prepared Bell states are not affected.
It is also worth mentioning that the extreme adiabatic limit is not necessary for achieving maximal entanglement. Nonadiabatic effects, i.e. $s < 1$, only result in a lower success probability $R$ as described by \eqref{eq:EntanglementRate}. 
This is mainly due to the fact that as a result of destructive interference the ring setup  annihilates unwanted effects originating from wave packet distortions of the $\sigma^+$ photon at the location of the second emitter $\EII$.
For these interference effects, however, it is important that effects of decoherence or dissipation between the wave packet components propagating along different paths inside the waveguide are suppressed and that quantum coherence is maintained.

The presented scheme for deterministic generation of Bell states in the extreme adiabatic limit may offer advantages in comparison with other probabilistic schemes \cite{Stephenson2020IonIon,Hucul2015AtomAtomEntanglement} by requiring only a single photon detection as a herald for entanglement instead of coincident two-photon measurements and by offering high success probabilities.  
As our scheme is rather insensitive to the 
pulse form of the photon wave packet it may also offer advantages over previously proposed 'send-receipt' schemes \cite{Rempe2012ElementaryNode}. 
As adiabatic excitation transfer 
has already been realized in a cavity scenario \cite{Rempe2020} and has led to entanglement generation schemes in this regime \cite{Rempe2021CNOTEntanglement} our proposal might be particularly interesting for applications involving waveguide geometries.

\section{Quantum Cloning} \label{sec:Cloning}

In this section it is shown that it is possible to copy 
an arbitrary qubit state of the polarization degrees of freedom of a single photon wave packet onto two quantum emitters by exploiting the indistinguishability of different pathways of the spontaneously emitted photon in adiabatic excitation transfer processes. Thereby it is possible to realize not only universal optimal symmetric quantum cloning processes, as discussed in Sec. \ref{sec:SymmetricCloning}, but also more general asymmetric cloning processes, as discussed in Sec. \ref{sec:CloningAsymmetric}.

\begin{figure}[t]
    \centering
    \includegraphics[width=\linewidth]{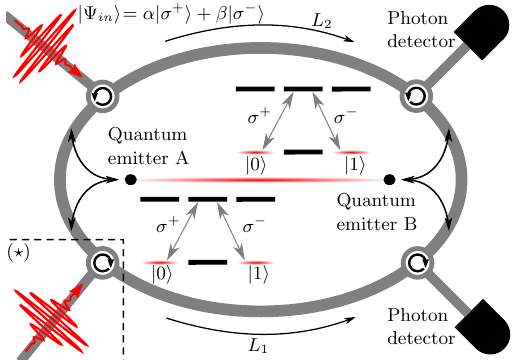}
	\caption{Quantum cloning is achieved with the same setup as for entanglement generation with the only difference of polarization-insensitive photon detectors. The quantum emitters are initially prepared in a Bell state.\\
	The part inside the dashed box ($\star$) is only present in the asymmetric cloning scheme. For asymmetric cloning the incident photon is split, e.g. by a beam splitter (not depicted), and injected via two paths into the waveguide ring. Detection of a photon at either photon detector heralds successful cloning.}
	\label{fig:CloningScheme}
\end{figure}

\subsection{Symmetric cloning of a photonic qubit onto two quantum emitters} \label{sec:SymmetricCloning}

We consider the same ring-shaped setup as described in Sec. \ref{sec:RingSetup} for entanglement generation under the same idealized assumptions. Thus, two distant quantum emitters $\EI$ and $\EII$ are interconnected by a waveguide ring as depicted in Fig.~\ref{fig:CloningScheme}. They are excited adiabatically by a single-photon wave packet in such a way that the coupling of this photon outside of the waveguide is negligible.
The only major difference is that now the two photon detectors are insensitive to the polarization of the spontaneously emitted photon emitted by the emitters, i.e. they are assumed to perform a von Neumann measurement without selecting a particular polarization component.

The scheme depicted in Fig.~\ref{fig:CloningScheme} aims at cloning an initially prepared pure qubit state involving an arbitrary linear superposition of a $\sigma^+$-polarized and a $\sigma^-$-polarized single photon wave packet injected into the ${\fa}$-direction of the ring-shaped waveguide with the help  of an optical circulator.
The two quantum emitters $\EI$ and $\EII$ are initialized in a Bell state. This preparation may be achieved with the help of the entanglement generation scheme discussed in Sec. \ref{sec:RingSetup}, for example. Using the notation of Sec. \ref{sec:RingSetup},
the initial quantum state of the emitter-field system is given by
\begin{align}
    \ket{\Psi(t_0)} =& \big( \alpha\ket{\sigma^+}^{\fa} + \beta\ket{\sigma^-}^{\fa} \big) \nonumber\\
    &\otimes \big( \ket{1}^\EI\ket{1}^\EII + \ket{0}^\EI\ket{0}^\EII \big) /\sqrt{2} \label{eq:CloningInitial}
\end{align}
with  $\abs{\alpha}^2+\abs{\beta}^2=1$. Assuming that the propagation properties inside the waveguide are polarization independent, 
at the location of emitter $\EI$
the one-photon amplitudes $f^A_+(t)$ and $f^A_-(t)$ associated with the normalized single-photon states
$\ket{\sigma^+}^{\fa}$ and $\ket{\sigma^-}^{\fa}$ are identical and fulfill the relations
$\int_{t_0}^{\infty} \abs{f^A_+(t)}^2dt = \int_{t_0}^{\infty} \abs{f^A_-(t)}^2dt = 1$.
As spontaneous decay
 out of the waveguide is assumed to be negligible the efficiency of adiabatic quantum state transfer at emitter $\EI$ is given by $\eta=1/2$ (cf. \eqref{eq:EntangelementEfficiency}).

The resulting  pure quantum state of the emitter-field system after two interactions with both emitters A and B can be determined by considering the components of the initial state \eqref{eq:CloningInitial} individually. The initial state component
$\ket{\sigma^+}^{\fa}\ket{0}^A\ket{0}^B$ 
has served as the initial state in Sec.~\ref{sec:RingSetup} and according to \eqref{eq:EntanglementRingFinal} it leads to an entangled Bell state between both emitters. In view of the symmetry of the excitation scheme the same consideration applies to the initial state component $\ket{\sigma^-}^{\fa}\ket{1}^A\ket{1}^B$. 
The other two components of the initial emitter-field quantum state in \eqref{eq:CloningInitial}, namely $\ket{\sigma^+}^{\fa}\ket{1}^A\ket{1}^B$ and $\ket{\sigma^-}^{\fa}\ket{0}^A\ket{0}^B$, will not cause an excitation transfer at emitter A or B. This is due to the fact that a $\sigma^+$ ($\sigma^-$)-polarized photon cannot excite quantum emitter A or B if it is initially prepared in the ground state $|1\rangle^\EI$ ($|0\rangle^\EI$) or $|1\rangle^\EII$ ($|0\rangle^\EII$).
Thus, after both interactions in the extreme adiabatic limit, i.e. $\Delta \omega/\Gamma \to 0$ implying $s\to 1$ and $r\to 0$,
the final quantum state of the emitter-field system is given by
\begin{align}
    \ket{\Psi_{\text{fin}}(t)} =   &\frac{\alpha}{\sqrt{2}}  \ket{\sigma^+}^{\fa}\ket{1}^\EI\ket{1}^\EII + \frac{\beta}{\sqrt{2}}\ket{\sigma^-}^{\fa}\ket{0}^\EI\ket{0}^\EII \nonumber\\
    +& \frac{\alpha}{\sqrt{8}} \Big(  \big[ \ket{\sigma^-_\EI(t)}^{\fc} + \ket{\sigma^-_\EI(t)}^{\fa} \big]\ket{1}^\EI\ket{0}^\EII \nonumber \\
    &\quad\: +  \big[ \ket{\sigma^-_\EII(t)}^{\fc} + \ket{\sigma^-_\EII(t)}^{\fa} \big]\ket{0}^\EI\ket{1}^\EII \Big) \nonumber \\
    +&\frac{\beta}{\sqrt{8}} \Big(  \big[ \ket{\sigma^+_\EI(t)}^{\fc} + \ket{\sigma^+_\EI(t)}^{\fa} \big]\ket{0}^\EI\ket{1}^\EII \nonumber \\
    &\quad\: +  \big[ \ket{\sigma^+_\EII(t)}^{\fc} + \ket{\sigma^+_\EII(t)}^{\fa} \big]\ket{1}^\EI\ket{0}^\EII \Big)
    \label{eq:cloning1}
\end{align}
with all photon states involved being normalized to unity. 
Corrections to this extreme adiabatic limit can be obtained in a straightforward way from the general relation  \eqref{eq:EntanglementRingFinal} on which this result is based.
After the completion of both interactions the photon acts as an ancilla for a symmetric cloning process. This becomes apparent by determining the reduced density operators of emitter $\EI$ or $\EII$. Thereby, two cases may be distinguished depending on whether the two orthogonal directions of propagation $\fc$ and $\fa$ are selected or not.

First of all let us consider the case where a photon is detected in the $W_{\fa}$ reservoir without selecting its polarization.
In the extreme adiabatic limit at the position of the photon detector $\textbf{x}_D$ the one-photon amplitudes of the photon states $\ket{\sigma^\pm_\EI(t)}^{\fa} = \ket{\sigma^\pm_\EII(t)}^{\fa}$  are indistinguishable. This is apparent from
 \eqref{eq:WefOnePhotonAmplitude}.
Furthermore, apart from a factor of $(-1)$ the
one-photon amplitudes of
the photon states
$\ket{\sigma^\pm(t)}^{\fa}$ are identical to the ones of \eqref{eq:PhotonAmplitudeAdiabatic} at this detector position. 
This factor of $(-1)$ may be absorbed in the correlated emitter state so that 
the orthonormal photon basis is mapped onto the orthonormal emitter basis according to the relations
$\ket{\sigma^+}^{\fa} \mapsto \ket{1}$ and $\ket{\sigma^-}^{\fa} \mapsto -\ket{0}$. As a result the initially prepared photon polarization state $\alpha \ket{\sigma^+}^{\fa}+ \beta \ket{\sigma^-}^{\fa}$ is mapped onto the ideal emitter reference state $\ket{\Psi_\text{ref}}^i = \alpha\ket{1}^i - \beta\ket{0}^i$ of emitter $i\in \{A,B\}$. This is apparent from the reduced density operator of emitter $\EI$, for example, resulting from this projective von Neumann measurement in the extreme adiabatic limit. Tracing out the degrees of freedom of the field ($F$) and of emitter $\EII$ yields the result
  \begin{eqnarray}
    \rho_\EI^{\fa}  &=&
    {\rm Tr}_{B,F}\left(P_{\fa}|\Psi_{\text{fin}}(t)\rangle \langle \Psi_{\text{fin}}(t)|\right)/ p^{\fa} \nonumber\\
     &=& \frac{1}{6} {\bf 1}_A  + \frac{2}{3}
     \ket{\Psi_\text{ref}}^{AA}\!\bra{\Psi_\text{ref}}.
    \end{eqnarray}
Thereby, the projection operator $P_{\fa} = \sum_{j=+,-} |\sigma^j\rangle^{\fa}\langle \sigma^j\vert^{\fa}$ characterizes the von Neumann measurement of the photon propagating in the $\fa$ direction without selection of polarization. An expression of the same form is obtained for emitter $B$.
These emitter states are optimal copies and in the extreme adiabatic limit are heralded with a photon detection probability of $p^{\fa} = \langle \Psi_{\text{fin}}(t)|P_{\fa}|\Psi_{\text{fin}}(t)\rangle = 3/4$. 
The quality of this cloning process can be measured by the
fidelity $F^{\fa}$ of emitter  $\EI$ or $\EII$, i.e.
$F^{\fa} = ~^A\bra{\Psi_\text{ref}} \rho_A^{\fa}\ket{\Psi_\text{ref}}^A =  ~^B\bra{\Psi_\text{ref}} \rho_B^{\fa}\ket{\Psi_\text{ref}}^B$. 
Including also non-adiabatic corrections this fidelity is given by 
\begin{align}
    F^{\fa} &=  \frac{2}{3}\big[1+(s^2+r^2)/4 + \abs{\alpha\beta}^2 (s^2+r^2 -1)\big] \nonumber \\
    &= \frac{5}{6} \left[1 - \frac{4}{5}(1+4\abs{\alpha\beta}^2)  \frac{\Delta\omega^2}{\Gamma^2} + O\left(\frac{\Delta\omega^4}{\Gamma^4}\right) \right]
\end{align}
with the corresponding detection probability
\begin{align}
    p^{\fa} &= \langle \Psi_{\text{fin}}(t)|P_{\fa}|\Psi_{\text{fin}}(t)\rangle \nonumber \\
    &= \frac{3}{4}  + \frac{2 \Delta\omega^2}{\Gamma^2}   + O\left(\frac{\Delta\omega^4}{\Gamma^4}\right).
\end{align}
This detection probability does not depend on $\alpha,\beta$, because the probabilities for excitation transfer and hence the optical paths of the photon are polarization-independent.
In the extreme adiabatic limit, the cloning process is universal, i.e. its fidelity $F^{\fa}$ is independent of $\alpha$ and $\beta$, and attains the well-known optimal value of $F^{\fa} = 5/6$ \cite{Buvzek1996UniversalCloningMachine}. In this case its success probability is given by $p^{\fa}=3/4$.

If a photon is detected in the $W_{\fc}$ reservoir, from \eqref{eq:cloning1} we easily deduce  
the final emitter states as $ \rho^{\fc}_\EI  = \frac{1}{2} {\bf 1}_{\EI}$ and $\rho^{\fc}_\EII  =  \frac{1}{2} {\bf 1}_{\EII}$ in the extreme adiabatic limit.
Hence, all information about the initially prepared photon state is lost. This process takes place with a photon
 detection probability of $p^{\fc}=1 - p^{\fa}$ and including non-adiabatic corrections the resulting fidelity reads 
\begin{equation}
    F^{\fc} = \frac{1}{2}  - \frac{2 \Delta\omega^2}{\Gamma^2}   + O\left(\frac{\Delta\omega^4}{\Gamma^4}\right).
\end{equation}
Alternatively, if the single photon is measured at either detector in a von Neumann measurement without selection of the measurement result, in each single shot of the photon detection process 
a copy is obtained with fidelity
\begin{equation}
    F_{\text{one-shot}} = p^{\fa}  F^{\fa} 
    + p^{\fc} F^{\fc} 
\end{equation}
Hence, 
in this case universal cloning is achievable with a fidelity of $3/4$ in the extreme adiabatic limit. This fidelity is still well above the classical threshold of $2/3$ \cite{Gisin2005CloneReview}.

The scheme discussed in this section intrinsically ensures that both output emitter states are equal.
In the subsequent section we consider a modification of the ring setup which breaks this symmetry and allows for more general cloning scenarios.

\subsection{Asymmetric cloning of a photonic qubit onto two quantum emitters} \label{sec:CloningAsymmetric}

In this section we investigate a variation of the previously considered situation as depicted in Fig. \ref{fig:CloningScheme} (including area labeled $\star$). A photon wave packet is split into two equal amplitudes by a polarization-independent balanced beam splitter, is injected into the waveguide  at two entry points and propagates towards quantum emitter $\EI$. 
The two quantum emitters $\EI$ and $\EII$ are prepared in a Bell state. This preparation can be achieved by the entanglement generation scheme of Sec. \ref{sec:Entanglement}, for example. Thus, the initial state of the photon-emitter quantum system  is given by
\begin{align}
    \ket{\Psi(t_0)} =\! \Big( \alpha &\big(\! \ket{\sigma^+}^{\fc} + \ket{\sigma^+}^{\fa} \big)  + \beta \big(\! \ket{\sigma^-}^{\fc} + \ket{\sigma^-}^{\fa} \big) \Big) / \sqrt{2} \nonumber \\
    &\:\otimes \: \big( \ket{1}^\EI\ket{0}^\EII + \ket{0}^\EI\ket{1}^\EII \big) / \sqrt{2} \label{eq:InitialAsymmetricCloning}   
\end{align}
with $\vert \alpha \vert^2+\vert\beta\vert^2 = 1$. It should be mentioned that the symmetric scheme of Sec. \ref{sec:SymmetricCloning} employed a different Bell state.

We assume that the propagation properties inside the waveguide are polarization independent so that at the location of emitter $\EI$ the normalized one-photon amplitudes $f^\EI_{+,{\fa}}(t)$, $f^\EI_{+,{\fc}}(t)$, $f^\EI_{-,{\fa}}(t)$, and  $f^\EI_{-,{\fc}}(t)$ associated with the four fractions of the photon wave packet in \eqref{eq:InitialAsymmetricCloning} are identical. Hence, by detection of an emitted photon it is not possible to determine its previous path. For this purpose the first emitter $\EI$ has to be placed in such a position that the constructive interference conditions $e^{i\omega(L'_1-L'_2)/c}=1$ and $\vert L'_1 - L'_2\vert \ll \Delta\omega/c$ are satisfied. Thereby, $L'_1, L'_2$ are the two path lengths from the balanced beam splitter to emitter $\EI$. These conditions are analogous to the ones discussed in Sec. \ref{sec:RingSetup} for the second emitter, and enable perfect excitation transfer at emitter $\EI$ in the extreme adiabatic limit.

Analogous to the cloning scheme of Sec. \ref{sec:SymmetricCloning} 
the interactions with each emitter can be described by a linear superposition of two excitations induced by the $\sigma^+$ and the $\sigma^-$ components. Correspondingly, for the $\sigma^+$ component 
the relevant mode reservoirs are $W_{es} = W_{\fa}^+ \cup W_{\fc}^+$, $W_{ef} = W_{\fa}^- \cup W_{\fc}^-$ and $B_{es} = B_{ef} = \emptyset$. 
Analogously, for the $\sigma^-$ component 
the relevant mode reservoirs are $W_{es} = W_{\fa}^- \cup W_{\fc}^-$, $W_{ef} = W_{\fa}^+ \cup W_{\fc}^+$ and $B_{es} = B_{ef} = \emptyset$.

In order to realize different cloning scenarios within this setup we include an additional orthogonal decay channel for each of the excited emitter states as discussed in \eqref{eq:EfficiencyError}. It is assumed that for each of the emitters such a  spontaneous decay channel produces a photon with a different frequency which is subsequently lost. Denoting the spontaneous decay rates of these additional decay channels by $\gamma_i ^\text{other}$ for $i\in \{\EI,\EII\}$ the efficiency for adiabatic excitation transfer by emitter $i$ is given by
\begin{align}
    \eta_i = \bar{\eta}^2_i, \quad \text{ with } \quad \bar{\eta}_i = \frac{4\gamma^W_i}{4\gamma^W_i+\gamma^\text{other}_i}.
    \label{etas}
\end{align} 
Thus, the efficiencies can be tuned by modifying these additional decay rates $\gamma_i ^\text{other}$ 
\cite{Goban2015ExperimentalWaveguideScenario, douglas2015WaveguideCouplingTuning} so that various excitation transfer probabilities within the interval $[0,1]$ can be realized for each emitter. 
In the following it is demonstrated that this way various cloning scenarios can be realized.

For the sake of simplicity in the following we concentrate on the extreme adiabatic limit, i.e. $\Delta\omega/\Gamma \to 0$ implying $s\to 1$, $r\to 0$. Lowest order corrections to this limit scale similarly as for the scheme discussed in
Sec. \ref{sec:SymmetricCloning}.
Hence, provided a photon arrives at emitter $\EI$ with unit probability the probabilities for excitation transfer at the first emitter are given by $p_{s\rightarrow f} = p_{0\rightarrow 1} = p_{1\rightarrow 0} = \bar{\eta}^2_\EI$. The probabilities for spontaneous photon emission by this emitter are given by $p^{W_{es}} = (1-\bar{\eta}_\EI)^2, p^{W_{ef}} = \bar{\eta}_\EI^2$ and $p^{B_{es}} = p^{B_{ef}} = 0 $  with respective choices of $es$ or $ef$ for each polarization component of the one-photon wave packet. Spontaneous photon emission via the orthogonal decay channel characterized by $\gamma_{\EI}^\text{other}$ takes place with probability $p_{\EI}^\text{other} = 2(1-\bar{\eta}_\EI)\bar{\eta}_\EI$. Analogous probabilities hold for emitter B with $\bar{\eta}_B$.

The interaction of the various fractions of the one-photon wave packet with both emitters
can be determined analogously to the procedure used in Sec. \ref{sec:RingSetup}. 
In particular, in the extreme adiabatic limit the photon amplitudes of all polarization components are not distorted by the excitation processes $[$cf. \eqref{eq:PhotonAmplitudeAdiabatic} and \eqref{eq:WefOnePhotonAmplitude}$]$. They only experience possible phase-shifts of magnitude $\pi$.
As a result after interactions with both emitters 
the final photon-emitter state  is given by
\begin{align}
        \vert\Psi_{\text{fin}}&(t)\rangle   \nonumber\\
        =&+ \frac{\alpha}{2} \bigg( (1-\bar{\eta}_\EI) \Big[ \ket{\sigma^+(t)}^{\fc} + \ket{\sigma^+(t)}^{\fa} \Big]\ket{0}^\EI\ket{1}^\EII \nonumber \\
        &-\big(\bar{\eta}_\EI(1-\bar{\eta}_\EII) + \bar{\eta}_\EII\big) \Big[ \ket{\sigma^-(t)}^{\fc} + \ket{\sigma^-(t)}^{\fa} \Big]\ket{1}^\EI\ket{1}^\EII \nonumber \\ 
        &+ \Big.\big( \bar{\eta}_\EI\bar{\eta}_\EII + (1-\bar{\eta}_\EII)\big) \Big[ \ket{\sigma^+(t)}^{\fc} + \ket{\sigma^+(t)}^{\fa} \Big]\ket{1}^\EI\ket{0}^\EII  \bigg) \nonumber \\
        &+ \frac{\beta}{2} \bigg( (1-\bar{\eta}_\EI) \Big[ \ket{\sigma^-(t)}^{\fc} + \ket{\sigma^-(t)}^{\fa} \Big]\ket{1}^\EI\ket{0}^\EII \nonumber \\
        &-\big( \bar{\eta}_\EI(1-\bar{\eta}_\EII) + \bar{\eta}_\EII\big) \Big[ \ket{\sigma^+(t)}^{\fc} + \ket{\sigma^+(t)}^{\fa} \Big]\ket{0}^\EI\ket{0}^\EII \Big. \nonumber \\
        &+ \Big.\big( \bar{\eta}_\EI\bar{\eta}_\EII + (1-\bar{\eta}_\EII)\big) \Big[ \ket{\sigma^-(t)}^{\fc} + \ket{\sigma^-(t)}^{\fa} \Big]\ket{0}^\EI\ket{1}^\EII  \bigg) \nonumber \\
        &+ C \ket{\Psi_\text{other}(t)} \label{finalresult}
\end{align}
with $C= \sqrt{\bar{\eta}_\EI(1-\bar{\eta}_\EI)+(1-\bar{\eta}_\EI)^2\bar{\eta}_\EII(1-\bar{\eta}_\EII)}$.
The normalized state $\ket{\Psi_\text{other}(t)}$ describes all events in which the photon was spontaneously emitted via one of the additional orthogonal decay channels characterized by $\gamma_i^\text{other}$ for $i\in \{\EI,\EII\}$. These events take place with probability $C^2$.
In the extreme adiabatic limit the field states $\ket{\sigma^\pm(t)}$ are equal to the initial states $\ket{\sigma^\pm}$. 

The quantum state of \eqref{finalresult} describes a general cloning process in which the one-photon wave packet acts as an ancilla and in which the orthogonal photonic polarization states are mapped onto emitter states according to the relation
 $\ket{\sigma^+} \mapsto \ket{1}$ and $\ket{\sigma^-} \mapsto -\ket{0}$. As a result the initially prepared pure photonic polarization state
 $\alpha\ket{\sigma^+} + \beta\ket{\sigma^-}$
 is mapped onto the emitter states
 $\ket{\Psi_\text{ref}}^i = \alpha\ket{1}^i - \beta\ket{0}^i$ with $i\in \{\EI,\EII\}$.
Similarly as in the previous section the resulting one-shot cloning process becomes apparent if the field degrees of freedom are averaged out from this pure photon-emitter quantum state.
However, higher and even optimal fidelities can be achieved by conditioning this final state on detection of the spontaneously emitted photon by one of the photon detectors as this conditioning excludes photon loss via the decay channels characterized by the decay rates $\gamma_i^\text{other}$ with $i\in \{\EI,\EII\}$. 
It is apparent from the structure of the final state \eqref{finalresult} that both photon detectors yield identical results
if a von Neumann measurement without selection of the polarization is performed. Such a measurement process is described by a projection operator of the form $P_{\fa} = \sum_{j=+,-} |\sigma^j\rangle^{\fa}\langle \sigma^j\vert^{\fa}$.
The resulting quantum state of emitter $\EI$ is given by
    \begin{align}
        &\rho_\EI^{\fc} = \rho_\EI^{\fa} = {\rm Tr}_{B,F}\left(P_{\fa}|\Psi_{\text{fin}}(t)\rangle \langle \Psi_{\text{fin}}(t)| \right) / p^{\fa} \nonumber \\ &= \frac{R\abs{\alpha^2}+(1-\bar{\eta}_\EI)^2(\abs{\beta^2}-\abs{\alpha^2})}{R} \ket{1}\bra{1} \\
        &+ \frac{R\abs{\beta^2}+(1-\bar{\eta}_\EI)^2(\abs{\alpha^2}-\abs{\beta^2})}{R} \ket{0}\bra{0} \nonumber \\
        &- 2 \frac{(\bar{\eta}_\EI+\bar{\eta}_\EII-\bar{\eta}_\EI\bar{\eta}_\EII)(1-\bar{\eta}_\EII+\bar{\eta}_\EI\bar{\eta}_\EII)}{R}\left(  \alpha\beta^{\ast}\ket{1}\bra{0} + h.c. \right).\nonumber
    \end{align}
The resulting quantum state of the second emitter is analogous. 
The success probabilities $p^{\fa}$ ($p^{\fc}$) of this von Neumann measurement for detecting a $\fa$-($\fc$-)photon are given by
$p^{\fa} = p^{\fc} = R/2$ with
\begin{align} \label{eq:R}
    R = 1 - \bar{\eta}_\EI(1-\bar{\eta}_\EI) - (1-\bar{\eta}_\EI)^2\bar{\eta}_\EII(1-\bar{\eta}_\EII).
\end{align} 
The fidelity $F_i =  \bra{\Psi_\text{ref}} \rho_i^{\fa} \ket{\Psi_\text{ref}} = \bra{\Psi_\text{ref}} \rho_i^{\fc} \ket{\Psi_\text{ref}}$ with which
an initially prepared qubit state is copied onto emitter
$i$ is given by
    \begin{align}
        F_\EI &= 1 - (1-\bar{\eta}_\EI)^2\frac{1- 8\abs{\alpha\beta}^2\bar{\eta}_\EII(1-\bar{\eta}_\EII)}{2R}, \label{eq:Fid1}\\
        F_\EII &= 1 - \frac{(1-\bar{\eta}_\EII+\bar{\eta}_\EI\bar{\eta}_\EII)^2 - 8\abs{\alpha\beta}^2\bar{\eta}_\EI(1-\bar{\eta}_\EI)(1-\bar{\eta}_\EII) }{2R}. \label{eq:Fid2}
    \end{align}
\begin{figure}
    \centering
    \includegraphics[width=\linewidth]{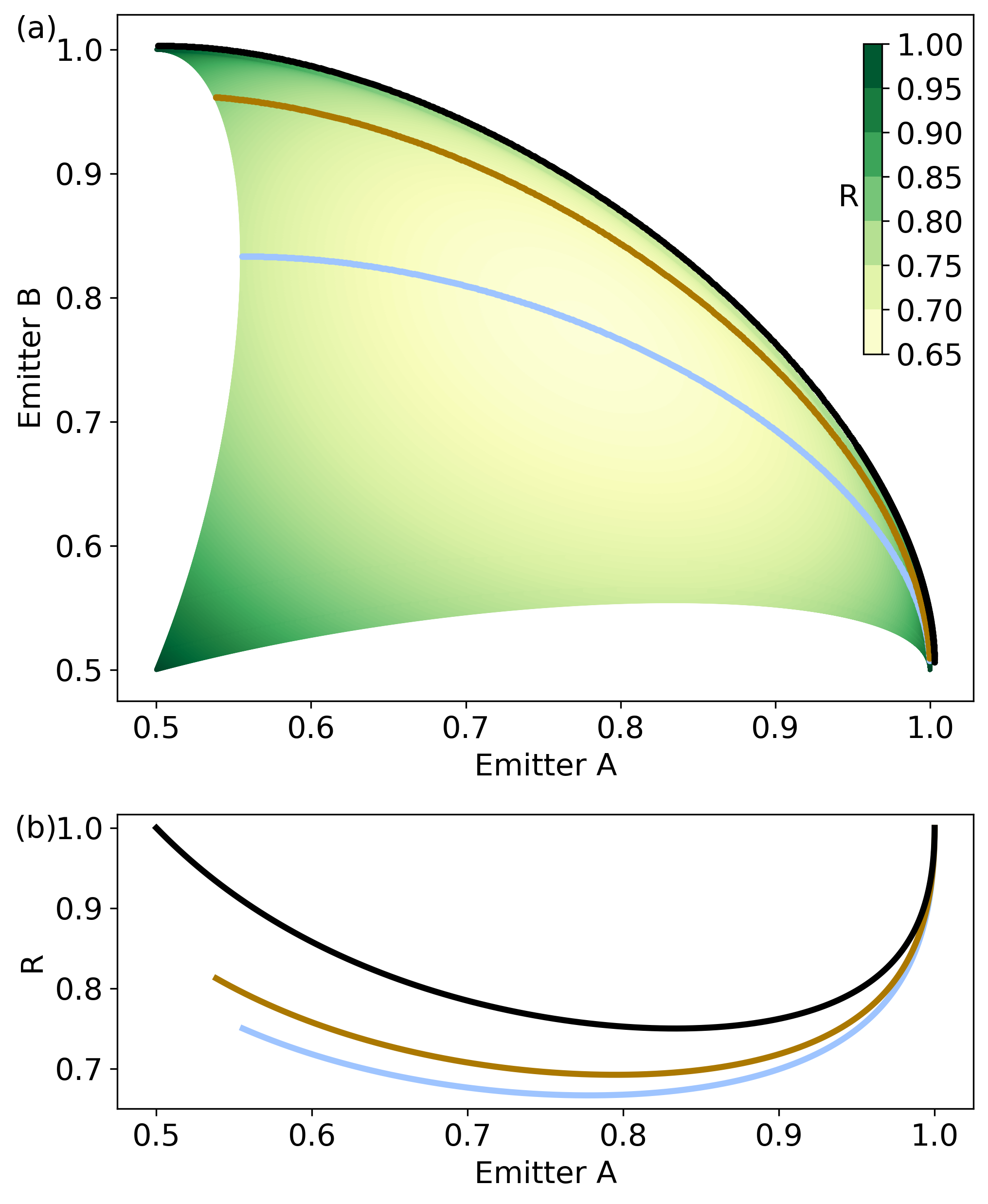}
    \caption{Success probabilities $R = p^{\protect\rotatebox[origin=c]{90}{\footnotesize $\circlearrowleft$}} + p^{\protect\rotatebox[origin=c]{270}{\footnotesize $\circlearrowright$}}$ $[$cf. \eqref{eq:R}$]$ 
    and achievable fidelities $F_A$ and $F_B$ $[$cf. \eqref{eq:Fid1}, \eqref{eq:Fid2}$]$ averaged uniformly over all input states on the Bloch sphere \cite{Buvzek1996UniversalCloningMachine, Gisin1997OptimalCloning}: (a) Each possible cloning process given by a pair of efficiencies $\bar{\eta}_A$ and $\bar{\eta}_B$ $[$cf. \eqref{etas}$]$ is plotted as a point with coordinates of the fidelities 
    $(F_{\EI},F_{\EII})$ of emitters $\EI$ and $\EII$. Its corresponding success probability $R$ is represented by color coding. Also shown are three lines of constant values of the efficiency of the second emitter, namely $\bar{\eta}_B = 1.0$ (black, upper line), $\bar{\eta}_B = 0.75$ (\textcolor{brown}{brown}, middle line), $\bar{\eta}_B = 0.5$ (\textcolor{grey}{grey}, lower line).
    (b) The success probability $R$ over the fidelity $F_\EI$ of emitter $\EI$ for these three lines of constant efficiencies $\bar{\eta}_B$.
    }
    \label{fig:fidelities}
\end{figure}
Therefore, within the full range of possible values of the efficiencies  $\bar{\eta}_\EI, \bar{\eta}_\EII \in [0,1]$ a variety of possible cloning processes can be realized. 
This is apparent from Fig.~\ref{fig:fidelities} in which for each pair of possible fidelities $(F_A,F_B)$ resulting from a pair of efficiencies $(\bar{\eta}_A,\bar{\eta}_B)$ the corresponding success probability $R = p^{\fa} + p^{\fc}$ for detection of a photon by one of the two photon detectors is depicted. Thereby, these fidelities are averaged over all possible pure input states on the Bloch sphere \cite{Buvzek1996UniversalCloningMachine, Gisin1997OptimalCloning}.
Also shown are three lines of constant efficiencies $\bar{\eta}_B$ together with the dependence of their success probabilities $R$ on the efficiency $\bar{\eta}_{\EI}$ of emitter $\EI$.
It is apparent from \eqref{eq:Fid1} and \eqref{eq:Fid2} that efficiencies fulfilling the relation 
$\bar{\eta}_\EII = \bar{\eta}_\EI/(1-\bar{\eta}_\EI)$ with $\bar{\eta}_\EI \leq \frac{1}{2} $ describe symmetric but not necessarily universal quantum cloning processes as in general their fidelities $F_A = F_B$ depend on the pure photonic input state.

Optimal symmetric or asymmetric cloning processes are described by 
the upper boundary of the surface in Fig.~\ref{fig:fidelities}. These cases are characterized by the absence of any additional orthogonal decay channel of the second emitter B, i.e. $\gamma_B^\text{other}=0$, so that $\bar{\eta}_\EII=1$ and perfect excitation transfer takes place at emitter $\EII$ in the extreme adiabatic limit. Under these conditions the fidelities become independent of the photonic input state. This optimal universal cloning is realized with fidelities 
\begin{align}
F_\EI^\text{opt} &= \frac{1+\bar{\eta}_\EI^2}{2(1-\bar{\eta}_\EI+\bar{\eta}_\EI^2)}, \:\:
F_\EII^\text{opt} = \frac{1+(1-\bar{\eta}_\EI)^2}{2(1-\bar{\eta}_\EI+\bar{\eta}_\EI^2)} .
\end{align}
In general it is asymmetric as $F_A \neq F_B$.
The corresponding success probability for detection of a photon by any of the two photon detectors is given by $R = 1-\bar{\eta}_\EI(1-\bar{\eta}_\EI)$.
The lowest success probability of all these optimal  cloning processes is achieved with a success probability $R=3/4$ for $\bar{\eta}_\EI = 1/2$. In this case a symmetric universal cloning process is realized with the well known optimal fidelities $F_\EI = F_\EII = 5/6$ for an optimal symmetric cloner \cite{Gisin1997OptimalCloning}. 
The same fidelities are achieved by the symmetric cloning scheme presented in Sec.~ \ref{sec:SymmetricCloning}.  

From \eqref{eq:Fid1} and \eqref{eq:Fid2} it is also apparent that
phase-covariant quantum cloning can be realized by this scheme \cite{Gisin2005CloneReview} with input states restricted to the equatorial line of the Bloch sphere, i.e. $\vert \alpha\vert~ =~ \vert \beta\vert =1/\sqrt{2}$. 
In this case the fidelities of  \eqref{eq:Fid1} and \eqref{eq:Fid2} become independent of the input state. Choosing $\bar{\eta}_\EII=(1+\bar{\eta}_\EI)^{-1}$ the fidelities become optimal, i.e.
\begin{align}
F_\EI^{pc} &= \frac{1}{2} + \frac{\bar{\eta}_\EI}{1+\bar{\eta}_\EI^2}, \quad\quad
F_\EII^{pc} = \frac{1}{1+\bar{\eta}_\EI^2}. 
\end{align}
Thus, in general this scheme achieves optimal universal asymmetric phase covariant quantum cloning \cite{Zhang2007OptimalFidelities}.
In the special case of $\bar{\eta}_\EI=\sqrt{2}-1$ this cloning process is symmetric with $F_A = F_B = (2+\sqrt{2})/4$ and with a success probability of 
$R = 4(3-2\sqrt{2}) \approx 69\%$.

\section{Conclusion}

We have generalized our recently proposed resonant dissipation-enabled adiabatic quantum state transfer processes \cite{at15}. These processes are capable
of transferring passively the quantum state of the polarization degrees of freedom of a single-photon wave packet to material quantum emitters. In this generalization we have particularly concentrated on the so far unexplored properties of the spontaneously emitted photon. In addition, we have taken into account non-adiabatic corrections describing physical situations in which the Fourier-limited bandwidth of the exciting single-photon wave packet is not vanishingly small in comparison with the relevant dissipative decay rates. 

It has been demonstrated that the degrees of freedom of the spontaneously emitted photon can be exploited for quantum information processing. For this purpose we have proposed schemes for realizing entanglement generation between distant quantum emitters and quantum state cloning of the polarization degrees of freedom of a single photon within the framework of waveguide scenarios. As these schemes are passive they do not require any external control of pulse shapes of the single-photon wave packet or active feedback beyond postselection by photon detection.
Although our theoretical investigations have concentrated on waveguide scenarios their results are expected to be relevant also for other scenarios as long as the processes involved are adiabatic so that the Fourier-limited bandwidth of the resonantly exciting single-photon wave packet involved is small in comparison with the relevant dissipative rates. As the investigated adiabatic processes exploit a balancing of the dissipative processes involved, for experimental realizations it is important to take into account properly all relevant radiative couplings of the quantum emitters to the relevant electromagnetic field modes which induce these dissipative processes.
Our investigations also suggest other quantum technological applications of these resonant adiabatic dissipation-enabled processes, such as the implementation of quantum logical gates. First steps forward in this direction have already been taken \cite{Chinese, Rempe2021CNOTEntanglement}.

\section*{Acknowledgments}

The authors acknowledge fruitful discussions with S. A. Moiseev, E. Fitzke
and M. Tippmann. This research was supported by the Deutsche
Forschungsgemeinschaft (DFG), SFB 1119–236615297.

\appendix
\section{Time evolution in the Weisskopf-Wigner approximation} \label{app:DecayEquationDerivation}

In this appendix the dynamics of the quantum emitter excited by a single photon is investigated. Within the framework of the Weisskopf-Wigner approximation \cite{weisskopf1930wigner}
the differential equation \eqref{eq:DecayRatePsiE} is derived for the excitation amplitude $\Psi_e(t)$ of the excited emitter state.

Starting from the Hamiltonian of   \eqref{eq:Hamiltonian} and the ansatz of   \eqref{eq:Ansatz} for the quantum state, a direct integration of the time dependent Schrödinger equation yields the results

\begin{widetext}

\begin{align}
    \ket{\Psi_{es}(t)}^{W_{es},B_{es}} &= \frac{i}{\hbar} \int_{t_0}^t e^{-i\omega_{es}(t'-t_0)} \textbf{d}_{es}^{\ast}\cdot \left( \hat{\textbf{E}}_{W_{es}}^-(\textbf{x}_A,t') + \hat{\textbf{E}}_{B_{es}}^-(\textbf{x}_A,t') \right) \ket{0}^{W_{es},B_{es}} \Psi_e(t') dt' \: + \:\ket{\psi_{in}}^{W_{es}}\ket{0}^{B_{es}}, \label{eq:PsiES}\\
\ket{\Psi_{ef}(t)}^{W_{ef},B_{ef}} &= \frac{i}{\hbar} \int_{t_0}^t e^{-i\omega_{ef}(t'-t_0)} \textbf{d}_{ef}^{\ast}\cdot  \left( \hat{\textbf{E}}_{W_{ef}}^-(\textbf{x}_A,t') + \hat{\textbf{E}}_{B_{ef}}^-(\textbf{x}_A,t') \right) \ket{0}^{W_{ef},B_{ef}} \Psi_e(t') dt'. \label{eq:PsiEF} 
\end{align}
Accordung to the ansatz of   \eqref{eq:Ansatz} these field states are not normalized. 
Correspondingly, the probability amplitude of the excited emitter state $\Psi_e(t)$ fulfills the integro-differential equation 
\begin{align}
\frac{d}{dt}\Psi_e(t) =& -\frac{1}{\hbar^2} \sum\limits_{j\in \lbrace es, ef \rbrace} \int_{t_0}^t dt' e^{i\omega_j(t-t')}~^{W_j,B_j}\bra{0} \left[ \textbf{d}_{ej}\cdot  \left( \hat{\textbf{E}}_{W_{j}}^+(\textbf{x}_A,t) + \hat{\textbf{E}}_{B_{j}}^+(\textbf{x}_A,t) \right) \right] \label{eq:DecayIntegroDifferentialEq} \\
&\left[ \textbf{d}_{ej}^{\ast}\cdot  \left( \hat{\textbf{E}}_{W_{j}}^-(\textbf{x}_A,t') + \hat{\textbf{E}}_{B_{j}}^-(\textbf{x}_A,t') \right) \right] \ket{0}^{W_j,B_j}   \Psi_e(t') \quad + \quad \frac{i}{\hbar} e^{i\omega_{es}(t-t_0)}~^{W_{es}} \bra{0} \textbf{d}_{es}\cdot \hat{\textbf{E}}_{W_{es}}^+(\textbf{x}_A,t') \ket{\psi_{in}}^{W_{es}}. \nonumber
\end{align}
\end{widetext}
The inhomogeneous term of this equation involving the one-photon amplitude
\begin{align*} 
     f_{in}(t) = \frac{e^{i\omega_{es}(t-t_0)}}{\hbar\sqrt{\gamma^W_{es}}}  \! ~^{W_{es}}\bra{0}\textbf{d}_{es}\! \cdot \hat{\textbf{E}}^+_{W_{es}}(\textbf{x}_A, t)  \ket{\psi_{in}}^{W_{es}}\!\!.
\end{align*}
describes the excitation by the incident single photon. Inserting the electric field operator of \eqref{eq:ElectricFieldOperator} into the integral on the right hand side of \eqref{eq:DecayIntegroDifferentialEq} and taking into account the orthogonality of the waveguide and background modes we obtain the result
\begin{align} \label{eq:AppendixA4}
    \frac{d}{dt}\Psi_e(t) &= -\frac{1}{\hbar^2} \sum\limits_{j\in \lbrace es, ef \rbrace} \int_{t_0}^t dt' \sum_{\omega\in W_j, B_j}
    e^{i(\omega_j-\omega)(t-t')}  \nonumber \\ &\!\!\!\!\!\!\!\!\times \frac{\hbar\omega}{2\epsilon_0} \sum_{\lambda \in  W_j, B_j }
    \abs{\textbf{d}_{ej}^{\ast}\cdot \textbf{u}_{\omega,\lambda}(\textbf{x}_A) }^2 \Psi_e(t') \: + \: i\sqrt{\gamma_{es}^W} f_{in}(t).
\end{align}
The integral on the right-hand side can be evaluated with the help of the Weisskopf-Wigner approximation \cite{weisskopf1930wigner}.
This approximation is valid as long as the time evolution of $\Psi_e(t')$ is governed by a significantly longer time scale than the time evolution of the integral kernel originating from the sum of all relevant modes which couple to the quantum emitter.
Thus, in this approximation \eqref{eq:AppendixA4} reduces to the final result
\begin{equation*} 
    \frac{d}{dt}\Psi_e(t) \: = \: -\frac{\gamma^W_{es} + \gamma^B_{es} + \gamma^W_{ef} + \gamma^B_{ef}}{2}  \Psi_e(t) \: + \: i \sqrt{\gamma^W_{es}} f_{in}(t)
\end{equation*}
with the spontaneous decay rates
\begin{equation}
    \gamma^R_j \: = \: \frac{2\pi}{\hbar}   
    \sum\limits_{(\omega, \lambda)\in  R_j } \frac{\hbar\omega}{2\epsilon_0}  \delta(\hbar\omega_j-\hbar\omega)\abs{\textbf{d}_j\cdot  \textbf{u}_{\omega_j,\lambda}(\textbf{x}_A)}^2.
\end{equation}
According to the golden rule \cite{FermisGoldenRule} $\gamma_j^R$ is the decay rate of the excited quantum emitter state $\ket{e}$ originating from spontaneous emission of a photon with frequency $\omega_j$ with $j\in\lbrace es, ef\rbrace$ into the mode reservoir $R\in\lbrace W, B\rbrace$.

\section{Adiabatic limit of the transfer probability} \label{app:IteratedPartialIntegration}

In this appendix it is shown that in the adiabatic limit the transfer probability of \eqref{eq:TransferProbabilityPsiE} can be written in the equivalent form of \eqref{eq:TransferProbTermsOfFin}.

The one-photon amplitude $f_{in}(t)$ of \eqref{eq:fin}, which characterizes the excitation of the quantum emitter at position $x_A$, vanishes initially at time $t_0$ and after the interaction, i.e. $f_{in}(t_0) = f_{in}(t\rightarrow\infty ) = 0$. Furthermore, it rises and diminishes slowly so that in the adiabatic limit also all its relevant higher order derivatives vanish initially and at the end of the interaction, i.e. $f_{in}^{(n)}(t_0) = f_{in}^{(n)}(t\rightarrow\infty ) = 0$ for all relevant $n\in\mathbb{N}_0$. Therefore, by iterated partial integration we obtain the asymptotic relations
\begin{align*}
    &\int_{t_0}^{\infty} \left| \sum\limits_{n=0}^{\infty} \left( \frac{-2}{\Gamma}\right)^n   f_{in}^{(n)}(t) \right|^2 dt \\
    &= \int_{t_0}^{\infty} \sum\limits_{n,m=0}^{\infty} \left( \frac{-2}{\Gamma}\right)^{n+m}  f_{in}^{(m)}(t) f_{in}^{\ast (n)}(t) dt\\
    &= \int_{t_0}^{\infty} \sum\limits_{n,m=0}^{\infty} \left( \frac{-2}{\Gamma}\right)^{n+m}  (-1)^m f_{in}(t) f_{in}^{\ast (n+m)}(t) dt\\
    &= \int_{t_0}^{\infty} \sum\limits_{u=0}^{\infty}\left( \frac{-2}{\Gamma}\right)^{u}\sum\limits_{m=0}^{u}   (-1)^m f_{in}(t) f_{in}^{\ast (u)}(t) dt\\
    &= \int_{t_0}^{\infty} \sum\limits_{u=0}^{\infty}\left( \frac{-2}{\Gamma}\right)^{2u} f_{in}(t) f_{in}^{\ast (2u)}(t) dt\\
    &= \int_{t_0}^{\infty} \sum\limits_{u=0}^{\infty}\left( \frac{4}{\Gamma^2}\right)^{u}(-1)^u f_{in}^{(u)}(t) f_{in}^{\ast (u)}(t) dt
\end{align*}
which yield the result of \eqref{eq:TransferProbTermsOfFin}.

\section{Scalar product relation of the photon state} \label{app:ScalarProductRelation}

In this appendix the scalar product relation \eqref{eq:EmittedIncidentOverlap} is derived between the field state $\ket{\Psi_{es}}^{W_{es},B_{es}}$ of \eqref{eq:PsiES} and the incident field state $\ket{\Psi_{in}}=\ket{\psi_{in}}^{W_{es}} \ket{0}^{B_{es}}$. 

With the help of the one-photon amplitude $f_{in}(t)$  of \eqref{eq:fin} long after the interaction, i.e. at $t \to \infty$, the scalar product between the states $\ket{\Psi_{in}}$ and $\ket{\Psi_{es}}^{W_{es},B_{es}}$ can be written in the form
\begin{align}
    \lim\limits_{t\rightarrow\infty} &\bra{\Psi_{in}} \Psi_{es}(t) \rangle^{W_{es},B_{es}} \:\: = \nonumber \\
    &= 1 + i \int_{t_0}^{\infty} \sqrt{\gamma^W_{es}} f^{\ast}_{in}(t')\Psi_e(t') dt' \nonumber \\
    &=1 - \frac{\gamma^W_{es}}{\Gamma/2} \int_{t_0}^{\infty}  f^{\ast}_{in}(t')  \sum\limits_{n=0}^{\infty} \left( \frac{-2}{\Gamma}\right)^{n}\!
  f_{in}^{(n)}(t)  dt' \label{eq:AppCRelation2} \nonumber \\
  &= 1 - \frac{\gamma^W_{es}}{\Gamma/2} (s+ir), 
\end{align}
yielding the relation of \eqref{eq:EmittedIncidentOverlap}.
Thereby, the result of \eqref{eq:PsiEFullAnalytic} was used for $\Psi_e(t)$ in the second line of the derivation.  
The last line of this derivation is a consequence of dividing the sum over $n$ into even and odd terms and performing iterated partial integrations as in Appendix \ref{app:IteratedPartialIntegration}. 
The contributions involving even values of $n$ constitute the quantity $s$ $[$cf.   \eqref{eq:TransferProbabilitySumS}$]$ and the contributions involving odd values of $n$ constitute the quantity $r$ $[$cf. \eqref{eq:SumR}$]$.

\end{document}